%% file: dapp_stats_paper.tex
\documentclass[12pt]{article}

\input mycommands

\usepackage[margin = 1in]{geometry}
\usepackage[caption=false]{subfig}
\usepackage{floatrow}
\usepackage{algorithm2e}
\usepackage{ulem}
\RestyleAlgo{boxruled}

\usepackage[title]{appendix}

\title{Analyzing second order stochasticity of neural spiking under stimuli-bundle exposure}
\author{
Chris Glynn\\ {\small University of New Hampshire} \and Surya T Tokdar\\ {\small Duke University} \and Azeem Zaman\\ {\small Harvard University} \and Valeria C Caruso\\ {\small University of Michigan} \and Jeffrey T Mohl\\ {\small Duke University} \and Shawn M Willett \\ {\small Duke University} \and Jennifer M Groh \\{\small Duke University}}
\date{}

\def\a{{\rm A}}
\def\b{{\rm B}}
\def\ab{{\rm AB}}

\def\se{{\rm SE}}

\newif\ifmark

\usepackage{natbib}

\begin{document}
\maketitle

\marktrue

\begin{abstract}
Conventional analysis of neuroscience data involves computing average neural activity over a group of trials and/or a period of time.  This approach may be particularly problematic when assessing the response patterns of neurons to more than one simultaneously presented stimulus.  In such cases, the brain must represent each individual component of the stimuli bundle, but trial-and-time-pooled averaging methods are fundamentally unequipped to address the means by which multi-item representation occurs.
We introduce and investigate a novel statistical analysis framework that relates the firing pattern of a single cell, exposed to a stimuli bundle, to the ensemble of its firing patterns under each constituent stimulus. Existing statistical tools focus on what may be called ``first order stochasticity'' in trial-to-trial variation in the form of unstructured noise around a fixed firing rate curve associated with a given stimulus.
Our analysis is based upon the theoretical premise that exposure to a stimuli bundle induces additional stochasticity in the cell's response pattern, in the form of a stochastically varying recombination of its single stimulus firing rate curves. We discuss challenges to statistical estimation of such ``second order stochasticity'' and address them with a novel dynamic admixture Poisson process (DAPP) model. DAPP is a hierarchical point process model that decomposes second order stochasticity into a Gaussian stochastic process and a random vector of interpretable features, and, facilitates borrowing of information on the latter across repeated trials through latent clustering. We present empirical evidence of the utility of the DAPP analysis with synthetic and real neural recordings.
\end{abstract}

\section{Introduction}
The brain is capable of encoding multiple objects presented simultaneously. 
But the neural computing behind this complex operation -- of great relevance to computational and cognitive neuroscience -- remains poorly understood. Presently lacking are statistical models and tools to quantify the relationship between an individual cell's response to a bundle of stimuli presented together, and the ensemble of its response patterns evoked when each stimulus is presented in isolation. We fill this gap with a novel statistical analysis framework, developed under the theory that a cell's response to a stimuli bundle is a stochastically varying, dynamic combination of its single stimulus response patterns. Such a theory allows the possibility that each item in the stimuli bundle dominates the cell's response pattern during distinct periods of time. We have recently presented evidence in favor of such an interpretation for auditory and visual stimuli \citep{caruso2018single}.

For simplicity, and also limited by available experimental data, we restrict this discussion to stimuli bundles consisting of two stimuli, each of which evokes detectable and separable response patterns from a neural cell. Neural activity in each experimental trial is measured as a spike train recorded over a common time horizon. We assume  repeated trials are available from each of the following three experimental conditions, A: ``exposure to a stimulus A alone'', B: ``exposure to a stimulus B alone'', and, AB: ``exposure to stimuli A and B together''.

Statistical analysis of spike train data typically assumes an underlying, stimulus-driven response curve from which a stochastic point pattern of spiking times is generated on each experimental trial \citep{gerstein1960approach}; see \cite{, kass2005statistical} and the references therein for a comprehensive overview. The response curve, taken as a function of time, is interpreted to give the potentially time-varying expected firing rates of the cell in response to the given stimulus. Variations of the spike train across multiple trials is considered ``random noise'' around this expected rate curve, realized in the form of a random point pattern. We refer to such variation as {\it first order stochasticity}. Statistical analyses under this framework usually proceed by aggregating spike trains across trials to improve accuracy in estimating the underlying response curve. We adopt this framework to estimate the expected firing rate curves $\lambda_\a(t)$ and $\lambda_\b(t)$ associated with, respectively, stimulus A and and stimulus B.

The same framework, however, may not apply to the case when both stimuli A and B are presented together, and the brain perceives them as distinct signals (perhaps revealed by behavioral response). To the brain, the stimuli are not fused together as a novel combined stimulus, but remain a stimuli bundle with each signal maintaining its individuality. It is conceivable that exposure to a stimuli bundle may induce a second type of stochasticity in the cell's response. Each trial under condition AB may involve its own distinct response curve that combines both $\lambda_\a$ and $\lambda_\b$, with the combination depending on unmeasured upstream or contemporaneous representation of the stimuli bundle by other cells. 

We refer to such random but structural variation across trials as {\it second order stochasticity}. We distinguish second order stochasticity from a broader umbrella term {\it trial-to-trial variation} often used in the literature \citep{kass2005statistical, ventura2005trial}. Our focus is on quantifying variability that is information-encoding and intrinsic to the cell in a given experimental condition, as opposed to noisy fluctuations that may be caused by factors extrinsic to the primary stimuli. Quantifying the precise nature of this variability is the central focus of our analysis.

\section{Statistical analysis of second order stochasticity} 
\subsection{The dynamic averaging model}
Our general approach is to describe second order stochasticity as dynamic averaging, in which the relative contributions of A-like and B-like response patterns can vary across time on multiple scales. Specifically, we describe the rate curve behind any specific AB trial, as a convex combination $\alpha(t)\lambda_\a(t)+ (1 - \alpha(t)) \lambda_\b(t)$, involving a possibly time varying weight curve $\alpha(t)$.  Second order stochasticity manifests when the entire weight curve varies stochastically across AB trials, either stably within a trial but variably across trials or variably across both trials, and time within trials. 


\floatsetup[figure]{style=plain,subcapbesideposition=center}
\begin{figure}[!tp]
\centering
\setcounter{subfigure}{0}
\sidesubfloat[Cell 1]{\includegraphics[trim={0 0 0 0cm},clip,height=3.5cm]{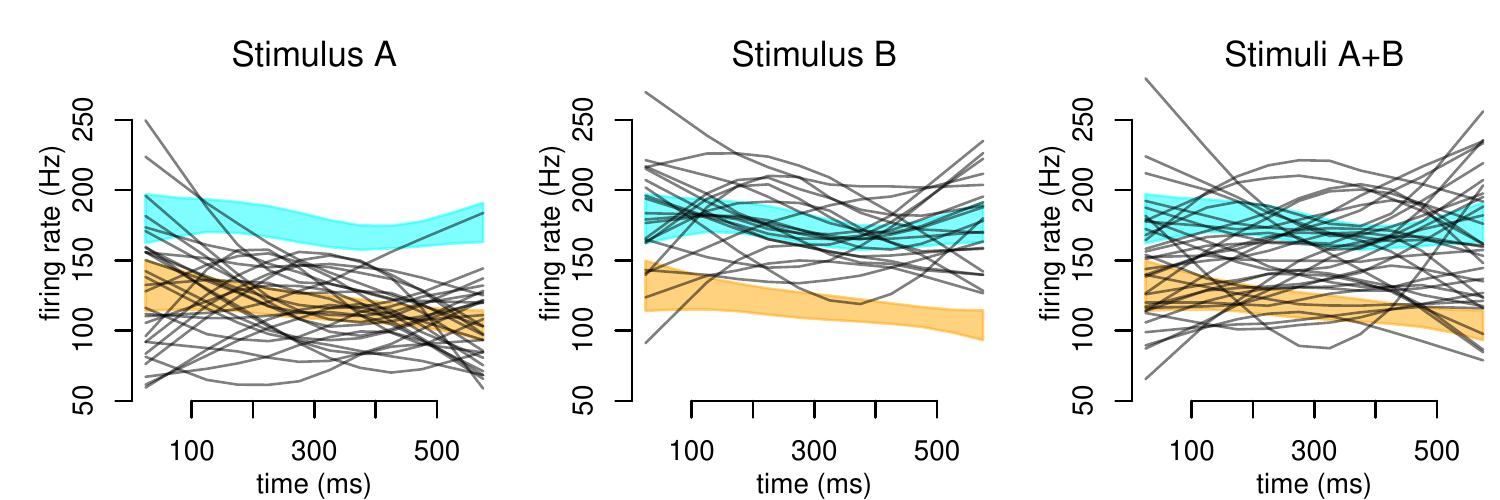}}\\
\sidesubfloat[Cell 2]{\includegraphics[trim={0 0 0 0cm},clip,height=3.5cm]{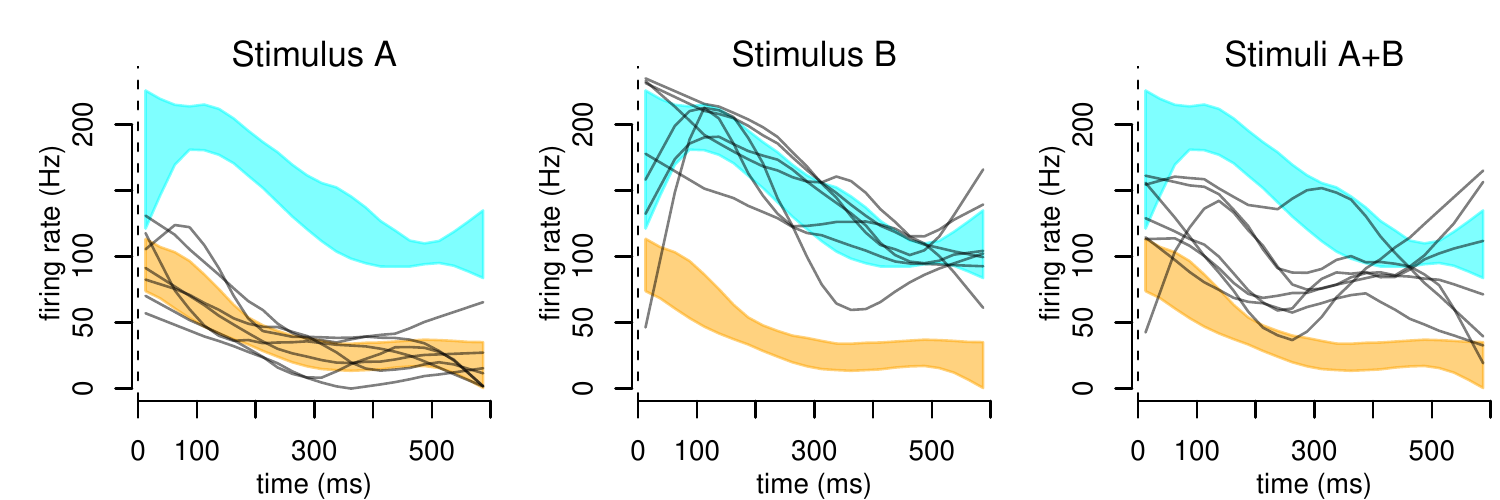}}\\
\sidesubfloat[Cell 3]{\includegraphics[trim={0 0 0 0cm},clip,height=3.5cm]{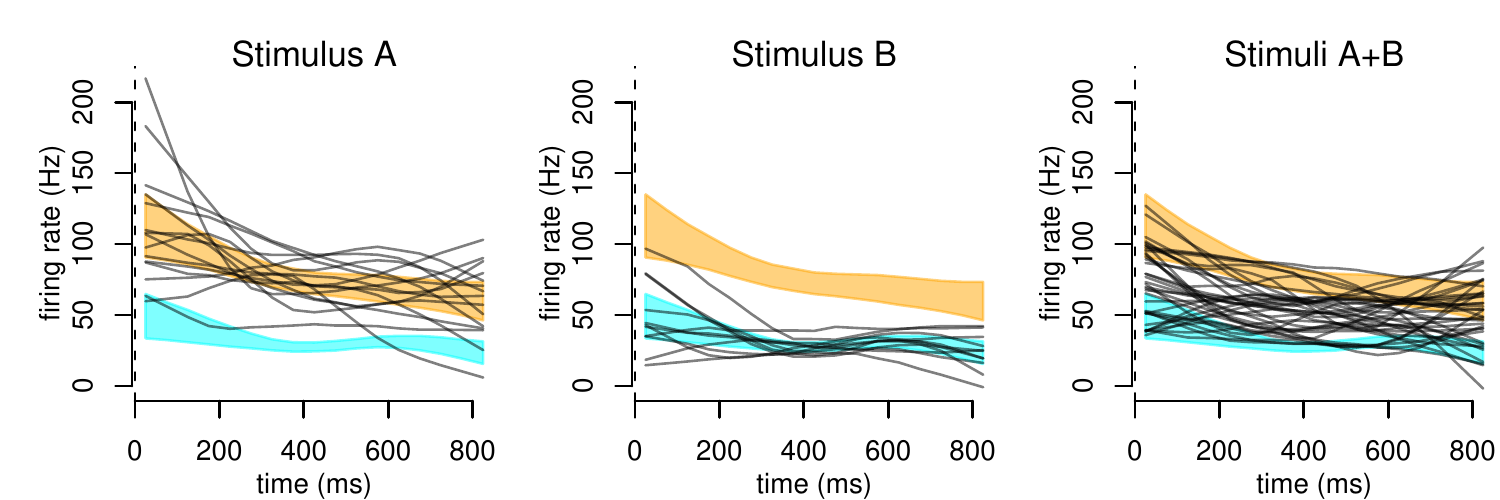}}
\caption{Multiple forms of stochasticity in inferior colliculus (IC). Each row corresponds to a distinct cell recorded from monkey IC and shows how the cell responds to the triple of experimental conditions A, B and AB, where A and B each corresponds to an auditory stimulus in the form of a bandpass filtered noise played from a certain angle. Each black curve represents one trial and shows the trial's spiking rate, which has been smoothed to aid visualization. The orange and cyan bands show estimates and uncertainty bands for $\lambda_A$ and $\lambda_B$. (a) Cell 1: AB responses appear to be a superimposition of A and B responses. (b) Cell 2: AB responses appear to fluctuate more widely within each trial than A or B responses. (c) Cell 3: similar to Cell 1 in appearance, but here the AB responses appear more squeezed toward the middle than how a superimposition of A and B responses would appear.} 
\label{fig 3 cells}
\efig

\begin{figure}[!tp]
\centering
\includegraphics[height=5cm]{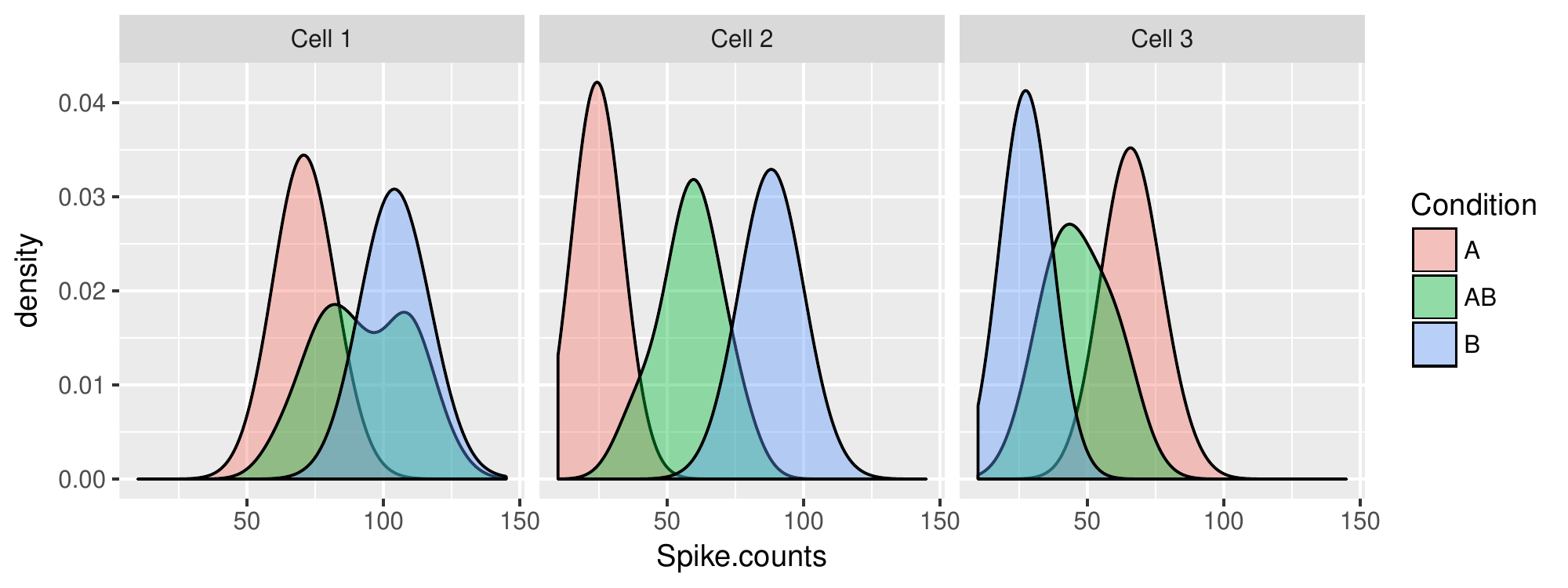}
\caption{Smoothed histograms of whole-trial spike counts of the three IC cells, grouped by experimental condition A, B and AB. For each cell, the AB total spike count distribution sits between the distributions under conditions A and B. But the shape of the AB distribution varies across cells.}
\label{reals-total}
\efig

A weight curve $\alpha(t)$ that is stable across time within a trial, but clusters bimodally near zero and one across trials, constitutes a special case, consistent with neurons encoding only one of the two stimuli per trial, and doing so in a fashion that is consistent with how they respond when only that stimulus is present.  In our previous study, we referred to such cases as showing whole-trial fluctuations \citep[``Mixtures'';][]{caruso2018single}. If the underlying firing rate dynamically alternates between those encoded by $\lambda_\a(t)$ and $\lambda_\b(t)$ within the course of a single trial, with $\alpha(t)$ approaching values of 0 and 1 for periods of time, the neuron may be encoding each stimulus separately during distinct temporal epochs of sub-trial durations. 

In either of these two special cases, the neuron is imagined to encode for only a single stimulus at any given time point. Our dynamic averaging model goes beyond such {\it one signal at a time} view, and allows for cases where the neuron's firing rate at any time point on a AB trial is truly intermediary between its A-level and B-level firing rates at the same time point. Here, the weight curve $\alpha(t)$ is seen as undulating, within or across trials, between a range of values that are bounded away from the extremes of zero or one. 

In Figures \ref{fig 3 cells} and \ref{reals-total}, we visualize response patterns of three example inferior colliculus cells from our experiments, where, the first two cells appear to exhibit, respectively, random selection of a particular stimulus at the whole trial level, and, interleaving of signals related to each stimulus across time within a trial.  In contrast, for Cell 3 in Figures \ref{fig 3 cells} and \ref{reals-total}, the response pattern appears to match truly intermediary averaging that varies stochastically possibly across both time and trials.


\subsection{Statistical estimation challenges}
It is challenging to carry out statistical analysis of second order stochasticity under the dynamic averaging assumption.
First, purely from a statistical accuracy perspective, estimation of the weight curves is difficult because one has access to only one spike train for each unknown weight curve. An ordinary aggregation across the AB trials no longer helps in combining information. Instead, one must rely on a hierarchical model that relates the weight curves to each other through a few meaningful features which are then estimated jointly from the pooled data. 

Second, on a more conceptual level, simply estimating the weight curves is not enough to draw inference on the exact nature of the cell's second order stochasticity. What is more relevant is to be able to predict how the cell is going to respond if new trials were carried out under the AB condition. While the weight curves associated with the new trials cannot be predicted exactly, one should be able to predict what features these weight curves are likely to possess. 

We address these challenges with a novel hierarchical point process model. We formulate the distribution of the weight curves as an unknown quantity to be estimated from the data. We reduce the estimation complexity of this problem by assuming the unknown stochasticity of the weight curves can be decomposed into a known Gaussian process distribution on smooth curves and an unknown probability distribution on a vector of a small number of meaningful summaries of the weight curves. The latter unknown probability distribution is conceived to be a discrete distribution, and, is assigned a Dirichlet \citep{ferguson1973bayesian} process prior to carry out a full Bayesian estimation. The discreteness assumption induces a (stochastic) clustering of the AB trials, facilitating borrowing of information across weight curves. Estimating the unknown distribution of the weight curves leads immediately to realistic prediction of the features of the weight curve in future trials.





\section{Dynamic admixture point process model}
\subsection{Poisson process formulation}
Let $n_\a$, $n_\b$ and $n_\ab$ give the numbers of trials under, respectively, conditions A, B and AB. Each trial produces a distinct spike train measurement. We assume that any neural spike train recorded over a given response window $[0,T]$ is a realization of a stochastic point process $(N(t): t\in[0,T])$ where $N(t)$ denotes the spike count between time zero and $t$, $0 \le t \le T$. For each condition $e \in \{\a, \b, \ab\}$ and each trial $j \in \{1, \ldots, n_e\}$, let $N^e_j(t)$ denote the corresponding point process.

For conceptual simplicity and analytical tractability we make a Poisson distributional assumption on these three sets of point processes.
%
\begin{enumerate}
\item $N^\a_j$, $j = 1,\ldots, n_\a$, are independent realizations of an inhomogeneous Poisson process with intensity function $\lambda_\a(t)$, $t \in [0,T]$;
\item $N^\b_j$, $j = 1, \ldots, n_\b$, are independent realizations of an inhomogeneous Poisson process with intensity function $\lambda_\b(t)$, $t \in [0,T]$;
\item $N^\ab_j$, $j = 1, \ldots, n_\ab$ are independently distributed inhomogeneous Poisson processes but with distinct intensity functions. The intensity function of the $j$-th such process is given by 
\[
\lambda_j(t) = \alpha_j(t) \lambda_\a(t) + \{1 - \alpha_j(t)\} \lambda_\b(t), ~ t \in [0,T]
\]
where $\alpha_j(t) \in [0,1]$, $t \in [0,T]$ is a possibly time varying weight curve.
\end{enumerate}
To incorporate second order stochasticity in our model, we assume the weight curves $\alpha_j(t)$, $j = 1, \ldots, n_\ab$, are independently distributed according to some unknown probability law $\bbP$ on the space of weight curves. This probability law may be understood as a characteristic of the neural cell when subjected to condition AB. Estimation of $\bbP$ is the key goal of our statistical analysis. We call this model the {\it dynamic admixture of Poisson process} (DAPP) model.

\subsection{Modeling the stochasticity of weight curves}

The space of weight curves is large and complex, and statistical estimation of an unknown probability law on this space is next to impossible without strong structural assumptions. Below we introduce a model for $\bbP$ where the unknown stochasticity of the weight curve is reduced to an unknown stochasticity of only a limited number of its features, namely, the curve's long term average value, maximum deviation from the average, and, the extent of waviness around the average. The remaining stochasticity is assumed to be governed by a known probability law, namely a modified Gaussian measure.  

\subsubsection{A Gaussian probability law for curves on $[0,T]$}
\label{sec:gp}
To be specific, for any $\ell > 0$, let $C^\se_\ell : [0,T]\times[0,T] \to (0, \infty)$ denote the so called squared exponential kernel with characteristic length-scale $\ell$, given by 
\[
C^\se_\ell(s, t) = \sigma_0^2 \exp\left\{-\frac{(s - t)^2}{2\ell^2}\right\}, ~~s,t \in [0,T],
\]
where $\sigma_0^2$ is a fixed scalar to be discussed later. For any scalars $\phi$ and $\psi > 0$, let $\gp(\phi, \psi C^\se_\ell)$ denote the probability law of a Gaussian process $(\eta(t): t \in [0,T])$ with mean and covariance functions
\beq
\expct [\eta(t)] \equiv \phi, ~~\cov[\eta(s), \eta(t)] = \psi C^\se_\ell(s, t),~~t, s \in [0,T]. 
\label{gp}
\eeq
It is well known that $\gp(\phi, \psi C^\se_\ell)$ defines a Gaussian measure on the space of smooth curves on $[0,T]$. 

\begin{rem}
\label{rem1}
Random curves generated from this measure are not exactly periodic, but are systematically wavy in the sense that the number of times such a curve crosses any fixed level is a random variable with a finite expectation. Indeed, the expected number of up-crossings\footnote{crossing from below; see \cite{adler2009random}, Chapter 11.} of level $\phi$ is  precisely $T/(2\pi\ell) \approx 0.16 \cdot T/\ell$. Therefore, a $\gp(\phi, \psi C^\se_\ell)$ law favors flat or wavy curves depending upon whether $\ell$ is, respectively, large or small. With $\ell = 160\%T$, one expects little waviness since the expected number of up-crossing is only a tenth, whereas, with $\ell = 4\%T$ one expects four up-crossings and hence considerable waviness.  
\end{rem}

\begin{rem}
\label{rem2}
Furthermore, for any random curve $\eta$ generated from $\gp(\phi, \psi C^\se_\ell)$, the scalar $\phi$ gives the expected value of the curve at any time point $t$ as well as the expected value of its long term average $\bar \eta := (1/T) \int_0^T \eta(t) dt$. If $\eta'$ were another curve generated from the same law and independently of $\eta$, then $\expct \{\eta(t) - \eta'(t)\}^2 = 2\psi\sigma_0^2$ at every $t \in [0,T]$, and, hence $\psi$ represents the range of the curve across repeated random generations. Both $\psi$ and $\ell$ play a role in controlling the within-trial deviation of $\eta(t)$ around its long term average $\bar \eta$. This deviation can be quantified as
\beq
\expct\left[ \frac{1}{T}\int_0^T (\eta(t) - \bar \eta)^2 dt \right] = \psi \sigma_0^2\left\{1 - T^{-2}\iint_{[0,T]^2} e^{-\frac{(s - t)^2}{2\ell^2}} dsdt\right\}.
\label{dev}
\eeq
The right hand side of \eqref{dev} is a monotonically decreasing function\footnote{given by $\psi\sigma_0^2\{1 - f(\ell/T)\}$ where $f(r) = 2 [\sqrt{2\pi} r \{\Phi(r^{-1}) - 0.5\} - r^2  \{1 - \exp(-0.5 r^{-2})\}]$, $r \ge 0$.} of $\ell/T$, going from a maximum value of $\psi \sigma_0^2$ at $\ell = 0$ to 0 as $\ell/T \to \infty$. For $\ell = 4\%T$, the right hand side of \eqref{dev} equals $0.9 \cdot \psi \sigma_0^2$, which means, the within-trial deviation is expected to be 90\% of what across-trial variance of the curve at any single time point. On the other hand, for $\ell = 160\%T$, the within-trial deviation equals $0.03 \cdot \psi\sigma_0^2$, i.e, only 3\% of the across-trial variance.  
\end{rem}

\subsubsection{A hierarchical Gaussian measure model for $\bbP$}
\label{s:bbP}
We model $\bbP$ as the probability law of a random weight curve $\alpha(t)$ generated by the following sequence of operations:
\begin{align}
\mbox{draw}~~(\phi, \psi, \ell) & \sim \bbQ,
\label{pq1}\\
\mbox{draw}~~\eta & \sim \gp(\phi, \psi C^\se_\ell), 
\label{pq2}\\
\mbox{set}~~\alpha(t) & = \frac{1}{1 + e^{-\eta(t)}},~~ t\in [0,T],
\label{pq3}
\end{align}
where $\bbQ$ is an unknown probability law on $(-\infty, \infty)\times (0, \infty) \times (0,\infty)$, to be estimated from data. 
Even without \eqref{pq1}, one could simply take \eqref{pq2}-\eqref{pq3} as a model for $\bbP$ where the only unknown quantities are the three scalars $\phi, \psi$ and $\ell$, which would render parameter estimation far easier. Therefore it is important to justify why we must include \eqref{pq1} in our model for $\bbP$.

Consider the case where a cell's second order stochasticity is close to 50-50 random selection; in nearly half of the AB trials $\alpha(t) \approx 0.9, t \in [0,T]$, while in the other half, $\alpha(t) \approx 0.1, t \in [0,T]$. Suppose our model for $\bbP$ were based of only \eqref{pq2}-\eqref{pq3} with the vector $(\phi, \psi, \ell)$ being the only unknown quantity. In light of the remarks in Section \ref{sec:gp}, one would estimate $\phi \approx 0$ and both $\psi$ and $\ell$ large. Hence, the estimated $\bbP$ will produce $\alpha(t)$ curves that are nearly flat across time, but, with no discernible concentration around either the 0.1 mark or the 0.9 mark. Therefore, while the estimated model will provide great fit to the observed data, it will completely fail to learn the true nature of the second order stochasticity. 

Inclusion of \eqref{pq1} in modeling $\bbP$ offers a much richer framework to learn various kinds of second order stochasticity. The vector $(\phi, \psi, \ell)$ in \eqref{pq2} exerts direct control on several broad features of the random weight curve $\alpha$ in \eqref{pq3}, e.g., its waviness, range, long term average and deviation around the long term average. The unknown probability measure $\bbQ$ of $(\phi, \psi, \ell)$ represents the unknown nature of stochasticity of these broad features. 



\relax

\section{Bayesian inference: prior specification}
\label{s:curves}
Although \eqref{pq1}-\eqref{pq3} offers a great reduction of complexity in statistical estimation of $\bbP$, estimating the remaining unknown quantity, the probability measure $\bbQ$, still remains a challenging problem. We adopt a Bayesian inference technique to estimate $\bbQ$ from data where a well chosen prior distribution on $\bbQ$ offers further structural simplification and regularization through latent clustering. 

\begin{rem}[Notation]
\label{rem3}
Below we use the generic expression $p(x|y)$ to understand the conditional distribution and/or the conditional probability density function (pdf) of one variable $x$ given another variable $y$. We use $\pois(\mu)$ to denote the Poisson distribution with mean $\mu$; $\bin(n,p)$ to denote the binomial distribution with size $n$ and success probability $p$; $\nm(m, v)$ to denote the (possibly multivariate) normal distribution with mean (vector) $m$ and variance (matrix) $v$; $\bet(a, b)$ to denote the beta distribution with shape parameters $a$ and $b$; $\dir(a_1, \ldots, a_k)$ to denote the $k$-dimensional Dirichlet distributions with shape parameters $a_i$, $i = 1, \ldots, k$; $\gam(r, s)$ to denote the gamma distribution with shape $r$ and rate $s$ (so that mean is $r/s$); $\textit{IG}(r, s)$ to denote the inverse-gamma distribution with shape $r$ and rate $s$; $\delta_x$ to denote the Dirac probability measure that assigns probability one to a single atom $x$; and, for an ordered, finite set $A = \{a_1, \ldots, a_k\}$ of size $k$ and a probability vector $\bfp = (p_1, \ldots, p_k)$, $P_{A,\bfp}$ to denote the discrete distribution $\sum_{i = 1}^k p_i \delta_{a_i}$ supported on $A$ that assigns probability $p_i$ to atom $a_i$, $i = 1, \ldots, k$.
\end{rem}

\subsection{A structural simplification of characteristic length-scale}
We restrict the characteristic length-scale $\ell$ to realize values in a known finite set $\scL = \{\ell^*_1, \ldots, \ell^*_L\} \subset (0, \infty)$. Such a choice offers great computational speed and can be justified by Remarks \ref{rem1} and \ref{rem2}. In particular, Remark \ref{rem1} implies that $\ell$ is intimately related to the number of stochastic oscillations of $\alpha$, with the expected number of up-crossing of its long-term average being $\approx 0.16 T / \ell$. Since this number can be limited to a finite range that is scientifically relevant, one could find a suitable finite set $\scL$ that offers a good coverage of plausible oscillatory behavior of the weight curves. For example, to represent between zero and four oscillations, one could work with $\scL = \{0.16 T / N: N \in \{0.1, 0.5, 1, 2, 3, 4\}\}$. In our experiments we typically have a response horizon of $T = 1000$ (measured in milliseconds), for which the corresponding grid, reordered from the smallest to the largest, is $\scL = \{40, 53.3, 80, 160, 320, 1600\}$.

We model $\bbQ$ hierarchically as the distribution of $(\phi, \psi, \ell)$ from the specification
\beq
(\phi, \psi, \bfpi) \sim \bbQ_{\phi, \psi, \bfpi}, ~~ \ell \sim P_{\scL,\bfpi}
\label{bfpi}
\eeq
where $\bfpi$ is a random element of the $L$-dimensional probability simplex $\Delta_L = \{(\pi_1, \ldots, \pi_L) \in \bbR^L: \pi_i \ge 0, \sum_i \pi_i = 1\}$, and, $\bbQ_{\phi, \psi, \bfpi}$ is an unknown probability measure on $(-\infty, \infty)\times(0,\infty)\times \Delta_L$. A prior distribution on $\bbQ$ is specified by assigning a prior distribution to $\bbQ_{\phi,\psi,\bfpi}$.


\subsection{Dirichlet process prior}
\label{dp prior}
We assign $\bbQ_{\phi,\psi,\bfpi}$ a Dirichlet process prior $\dpp(\kappa G_\kappa)$ with precision $\kappa > 0$ and base probability measure $G_\kappa$ on $(-\infty,\infty)\times(0,\infty)\times \Delta_L$ that depends on the precision, to be specified below. This prior specification restricts $\bbQ_{\phi, \psi,\bfpi}$ to be a (random) discrete probability measure with infinitely many atoms
\beq
\bbQ_{\phi, \psi,\bfpi} = \sum_{h = 1}^\infty \omega_h \delta_{(\phi^*_h, \psi^*_h, \bfpi^*_h)}
\label{dapp dp}
\eeq
where the atoms $(\phi_h^*, \psi^*_h, \bfpi^*_h)$, $h = 1,2,\ldots$, are drawn independently from the base measure $G_\kappa$ and the weights $\omega_h$, $h = 1,2,\ldots$, admit the stick-breaking representation $\omega_h = \beta_h\prod_{j = 1}^{h-1} (1 - \beta_j)$ with $\beta_h$, $h = 1,2,\ldots$, drawn independently from a $\bet(1, \kappa)$ distribution \citep{sethuraman1994constructive}. 

The discreteness of $\bbQ_{\phi,\psi,\bfpi}$ implies that repeated independent draws of $(\phi, \psi,\bfpi)$ from this probability law will produce duplication. Consequently, the AB trials can be grouped into clusters where within each cluster all trials have weight curves arising as in \eqref{pq2}-\eqref{pq3} with a single underlying $(\phi, \psi)$ and distinct realizations of $\ell$ arising from a shared probability vector $\bfpi$. Therefore these weight curves would have broad features such as the long term average and the range roughly matched. If $\bfpi$ was peaked in one coordinate, i.e, $\pi_i \approx 1$ for some $i$ while other $\pi_{i'}$ are close to zero, then the weight curves from the cluster would also have similar oscillatory behavior. However, in spite of sharing these broad features, the exact forms of these curves will be different.

The precision parameter $\kappa$ determines the extent of this clustering with larger values of $\kappa$ leading to more distinct clusters. Following common practice \citep{escobar1995bayesian} we assign the precision parameter a further $\gam(1,1)$ prior which makes the learning of this parameter relatively straightforward.

\begin{rem}
\label{rem4}
One could specify a Dirichlet process prior directly on $\bbQ$, without the introduction of the intermediary quantity $\bfpi$ as in \eqref{bfpi}. But our choice of  decoupling $\ell$ from $(\phi, \psi)$ via the introduction of $\bfpi$ leads to much improved posterior computation. See Appendix B for more details. 
\end{rem}

\subsection{An unconventional choice of the base distribution}
\label{base m}

We deviate from common practice in choosing the base measure $G_\kappa$ which we take to depend on the precision $\kappa$ and equal the law of $(\phi^*, \psi^*, \bfpi^*)$ where
\beq
\bfpi^* \sim \dir(a_1,\ldots, a_L),~~\psi^* |  \kappa \sim \bet(1, \kappa), ~~ \phi^* | (\psi^*, \kappa) \sim \nm(0, \sigma_0^2(1 - \psi^*)).
\label{g0}
\eeq
This choice of $G_\kappa$ ensures that under \eqref{pq1}-\eqref{pq2}, $\eta(t) \sim \nm(0, \sigma_0^2)$ at each $t \in [0,T]$. Consequently, with $\sigma_0 = 1.87$, our {\it a priori} belief is that $\alpha(t)$ is nearly uniformly distributed over the range $(0,1)$ at each single time point $t$. In contrast, the more conventional choice of a normal-inverse gamma base measure \citep{escobar1995bayesian} would lead to a heavy tailed Student-t prior on $\eta(t)$, and consequently, the prior on $\alpha(t)$ would place more mass than a uniform prior near the extremes of $\alpha(t) = 0$ and $\alpha(t) = 1$.   

The particular form of dependence of $G_\kappa$ on $\kappa$ in \eqref{g0} offers additional structural control on the clusters formed by repeated draws of $\eta$ from \eqref{pq1}-\eqref{pq2}. Let $\eta_j$, $j = 1, \ldots, n$, denote such repeated draws and focus on the behavior of $Y_j = \eta_j(t)$, $j = 1,\ldots,n$, at any arbitrary $t \in [0,T]$. We know that the marginal distribution of $Y_j$'s is $\nm(0, \sigma_0^2)$. This marginal distribution decomposes into a weighted average of the cluster specific distributions of $Y_j$'s with the weights being proportional to cluster sizes. When $\kappa$ is small, there is likely to be one dominating cluster of $\eta_j$'s sharing a common atom $(\phi^*, \psi^*)$. It is desirable that the $Y_j$'s in this dominating cluster should have a marginal distribution close to $\nm(0, \sigma_0^2)$. This is indeed the case under \eqref{g0}, which, for a small value of $\kappa$, makes $\psi^*$ likely to be close to 1, and, hence $\phi^*$ likely to be close to 0. 

On the other hand, for a large value of $\kappa$, there will be many clusters and it is desirable that the cluster specific distributions of $Y_j$'s should be distinct from each other. In this case for any cluster $\psi^*$ is likely to be small and $\phi^*$ is possibly away from zero, and, thus the corresponding marginal distribution of $Y_j$'s would be a normal distribution with a small variance and a center that is drawn randomly from a normal distribution with variance close to $\sigma_0^2$.

Although our nonconventional choice of $G_\kappa$ introduces some new computing challenges, they can be met with a fairly straightforward application of the widely used Algorithm 8 of \citet{neal2000markov}. The additional dependence of $G_\kappa$ on $\kappa$ in \eqref{g0} poses no serious obstacle to the learning of this precision parameter.

The hyperparameters $a_1, \ldots, a_L \in (0, \infty)$, of the Dirichlet distribution in \eqref{g0} determine the prior expectation for $\bfpi^*$ in the form of the probability vector $(a_1, \ldots, a_L)/\sum_i a_i$, with $\sum_i a_i$, called precision, serving as a measure of tightness of the prior around the prior expectation. For the default choice of $\scL$ as given before, and arranged from the smallest to the largest, we choose $a_i \propto i$ and adjust them so that $\sum_i a_i = 2$. With this choice, larger length-scales and hence flatter weight curves are slightly favored {\it a priori}. The precision value 2 ensures the prior belief to be at par with the information content of two observations drawn from the multinomial distribution $P_{\scL, \bfpi}$.


\section{Posterior computing}
\subsection{Time discretized model approximation}
For any step function $f(t)$ on $[0,T]$ that is continuous from the right, let $J(f) = \{t \in [0,T]: f(t) \ne f(t-)\}$ denote the set of its jump points. If $(N(t): t \in [0,T])$ is an inhomogeneous Poisson process with intensity $\lambda(t)$, then with probability one, $N$ is a step function that is continuous from the right and $J(N)$ is finite. In fact, the likelihood of observing $N$ can be expressed as 
\beq
\label{exact}
p(N|\lambda) = e^{-\int_0^T \lambda(t) dt} \prod_{t \in J(N)} \lambda(t)
\eeq
and may be used in a Bayesian update of a prior measure $\Pi$ on $\lambda$ to the posterior measure $\Pi(d\lambda|N) \propto p(N|\lambda) \Pi(d\lambda)$. 

However, since no closed form analytical expression is typically available for the posterior measure, one needs to employ numerical algorithms, e.g., Markov chain Monte Carlo (MCMC), to carry out posterior inference on $\lambda$. For such numerical algorithms, a direct use of this exact likelihood function creates serious computational challenges. The evaluation of the integral $\int_0^T\lambda(t)dt$ involves the entire curve $\lambda(t)$, $t \in [0,T]$. Consequently, the numerical algorithm needs to run on the infinite dimensional space of curves, presenting nearly insurmountable computational difficulties. \citet{rao2011gaussian} circumvent this problem by augmenting additional latent variables which allow them to run a Gibbs sampler for MCMC computation. While this technique could be directly implemented to draw posterior inference on $\lambda_\a$ (or $\lambda_\b$) based on only the A (B) trials data, its use in drawing inference on the $\alpha_j$ curves from AB trials data remains extremely challenging.

A less elegant but pragmatic alternative is to use time discretization. Fix an integer $M$ and partition the response window $[0,T]$ into $M$ contiguous subintervals $(0,w]$, $(w, 2w], \ldots, (T-w, T]$ of length $w = T/M$ each. Let $t^*_m = (m - 0.5)w$ be the midpoint of the $m$-th subinterval. When $M$ is relatively large, one can appeal to the Riemann approximation of $\int_0^T \lambda(t) dt$ and express \eqref{exact} as
\beq
p(N|\lambda) \approx \exp\left\{-\sum_{m = 1}^M w\lambda(t^*_m)\right\}\prod_{m = 1}^M \lambda(t^*_m)^{X_m} \propto \prod_{m = 1}^M \pois(X_m | w \lambda(t^*_m))
\label{approx}
\eeq
where $X_m = N(mw) - N((m-1)w)$ denotes the number of jumps in the $m$-th subinterval, 
and, the second and third terms are proportional as functions of $\lambda$.

By using \eqref{approx}, an MCMC now needs to be run only on the $M$-dimensional vector $(\lambda(t^*_1), \ldots, \lambda(t^*_M))$. Although one could obtain more accurate, $M$-term numerical approximation to $\int_0^T \lambda(t)dt$ by using Gaussian quadrature or Romberg's method, the equivalence of the second and third terms in \eqref{approx} is a real advantage of using the Riemann approximation as it allows us to develop an extremely efficient Gibbs sampler based MCMC algorithm for joint posterior inference on all model parameters.

\relax

\subsection{Reduced data and two-stage analysis}
Following the notation of the above subsection, let $X^e_{jm}$, denote the spike count in the $m$-th subinterval for the $j$-th trial under experimental condition $e$, where, $m = 1, \ldots, M$, $j = 1, \ldots, n_e$, $e \in \{\a, \b, \ab\}$. Under the approximation given by \eqref{approx}, our data model now looks as follows:
\benum
\item $X^\a_{jm} \sim \pois(w\lambda_\a(t^*_m))$, $m = 1, \ldots, M$, $j = 1, \ldots, n_\a$,
\item $X^\b_{jm} \sim \pois(w\lambda_\b(t^*_m))$, $m = 1, \ldots, M$, $j = 1, \ldots, n_\b$,
\item $X^\ab_{jm} \sim \pois(w\{\alpha_j(t^*_m)\lambda_\a(t^*_m) + (1 - \alpha_j(t^*_m))\lambda_\b(t^*_m)\})$, $m = 1, \ldots, M$, $j = 1, \ldots, n_\ab$,
\eenum
and all these random variables are independent of each other given $\lambda_\a$, $\lambda_\b$ and $\alpha_j$, $j = 1, \ldots, n_\ab$. Let $\bfX^e = (X^e_{jm} : 1 \le j \le n_e, 1\le m \le M)$ denote the $n_e \times M$ dimensional data matrix of bin counts from experiment $e \in \{\a, \b, \ab\}$.

Notice that only the AB trial data $\bfX^\ab$ is relevant to second order stochasticity analysis as it provides information on the $\alpha_j$ curves and their unknown feature generating distribution $\bbQ$. Below, we first describe how posterior inference can be drawn on these quantities from $\bfX^\ab$ alone under the working assumption that  $\lambda_\a$ and $\lambda_\b$ have already been estimated. Then, in Section \ref{s:first} we describe how the estimates of $\lambda_\a$ and $\lambda_\b$ may be obtained by analyzing $\bfX^\a$ and $\bfX^\b$ in a preprocessing step. We also discuss how the uncertainty in these estimates may be incorporated in the the second stage analysis of $\bfX^\ab$.

\subsubsection{MCMC inference for $\bbQ$ and $\alpha_j$ curves}
\label{s:mcmc}

Recall that underlying each $\alpha_j$ curve are a vector $(\phi_j, \psi_j, \bfpi_j) \sim \bbQ_{\phi,\psi,\bfpi}$, a scalar $\ell_j \sim \bfpi_j$ and a curve $\eta_j \sim \gp(\phi_j, \psi_j C^\se_{\ell_j})$ such that $\alpha_j(t) = [1 + \exp\{-\eta_j(t)\}]^{-1}$, $t \in [0,T]$, $j = 1, \ldots, n_\ab$. Clearly we can focus on the posterior distribution of these $\eta_j$ curves instead of the original $\alpha_j$'s. The other model parameters to be estimated are $\bbQ_{\phi,\psi,\bfpi}$ and the precision parameter $\kappa$. However,
\begin{align*}
p\big(\bbQ_{\phi,\psi,\bfpi}, \kappa, &~ \{\eta_j, \phi_j, \psi_j, \bfpi_j, \ell_j\}_{j = 1}^{n_\ab} ~\big|~ \bfX^\ab, \lambda_\a, \lambda_\b\big)\\
& \propto p\big (\kappa, \{\bfeta_j, \phi_j, \psi_j, \bfpi_j, \ell_j\}_{j = 1}^{n_\ab}~\big|~ \bfX^\ab, \lambda_\a, \lambda_\b\big)\\
& ~~~~~~~~ \times \prod_{j = 1}^{n_\ab} p(\eta_j | \bfeta_j, \phi_j, \psi_j, \ell_j) \\
& ~~~~~~~~
~~~~\times p(\bbQ_{\phi,\psi,\bfpi} | \kappa, \{\phi_j, \psi_j,\bfpi_j\}_{j = 1}^{n_\ab})
\end{align*}
where $\bfeta_j = (\eta_j(t^*_1), \ldots, \eta_j(t^*_M))$, $j = 1, \ldots, n_\ab$. Notice that each of the conditional probability distributions appearing in the last two lines above is available in closed form. Therefore, to obtain MCMC inference on all model parameters it suffices to focus on building a Markov chain sampler with target stationary distribution $p(\kappa, \{\bfeta_j, \phi_j, \psi_j, \bfpi_j, \ell_j\}_{j = 1}^{n_\ab} ~| \bfX^\ab, \lambda_\a, \lambda_\b)$.

Toward this goal, first rewrite our Poisson observational model $X^\ab_{jm} \sim \pois(w\{\alpha_j(t^*_m)\lambda_\a(t^*_m) + (1 - \alpha_j(t^*_m))\lambda_\b(t^*_m)\})$ as
\begin{align*}
(Z^\a_{jm}, Z^\b_{jm}) & \sim \pois(w \lambda_\a(t^*_m)) \times \pois(w\lambda_\b(t^*_m))\\
(Y^\a_{jm}, Y^\b_{jm}) & \sim \bin(Z^\a_{jm}, \alpha_j(t^*_m)) \times \bin(Z^\b_{jm}, 1 - \alpha_j(t^*_m))\\
X^\ab_{jm} & = Y^\a_{jm} + Y^\b_{jm},
\end{align*}
with independence assumed across $j = 1, \ldots, n_\ab$, $m = 1, \ldots, M$. This representation is valid since 
\beq
Z \sim \pois(\mu),Y | Z \sim \bin(Z, p) \implies (Y, Z - Y) \sim \pois(p\mu) \times \pois((1-p)\mu).
\label{poibin}
\eeq
Consequently, it is sufficient to construct a Markov chain sampler for the augmented target distribution 
\[
p(\bfZ^\a, \bfZ^\b, \bfY^\a, \bfY^\b, \kappa, \{\bfeta_j, \phi_j, \psi_j, \bfpi_j, \ell_j\}_{j = 1}^{n_\ab}~ |~ \bfX^\ab, \lambda_\a, \lambda\b)
\]
where $\bfZ^e = (Z^e_{jm}: 1\le j \le n_\ab, 1 \le m \le M)$, $\bfY^e = (Y^e_{jm}: 1\le j \le n_\ab, 1 \le m \le M)$, $e \in \{\a, \b\}$. Algorithm \ref{algo1} gives a schematic representation of our Markov chain sampler. All technical details are provided in Appendix \ref{app-mcmc}.

\begin{algorithm}[!t]
\KwIn{Binned spike counts $\bfX^\ab$ from AB trials, and, $\lambda_\a$ and $\lambda_\b$ curves (evaluated at the bin midpoints). Also, starting values for the model parameters $\kappa, \{\bfeta_j, \phi_j, \psi_j, \bfpi_j, \ell_j\}_{j = 1}^{n_\ab}$. These values maybe drawn from the prior.}
\KwOut{$S$ Markov chain samples of model parameters $\kappa, \{\bfeta_j, \phi_j, \psi_j, \bfpi_j, \ell_j\}_{j = 1}^{n_\ab}$}
\For{$s \gets 1$ \textbf{to} $S$} {
\parbox{0.8\textwidth}{
\benum
\item Impute $(\bfZ^\a, \bfZ^\b, \bfY^\a, \bfY^\b )$ by a combination of Poisson and binomial draws leveraging upon \eqref{poibin}.
\item Carry out a parameter-expanded Gibbs update of $\{\bfeta_j, \ell_j\}_{j = 1}^{n_\ab}$ by using the P\'olya-Gamma augmentation method of \citet{polson2013bayesian}.
\item Carry out a parameter-expanded Gibbs update of $\{\phi_j, \psi_j, \bfpi_j\}_{j = 1}^{n_\ab}$ by using Algorithm 8 of \citet{neal2000markov}.
\item Carry out a parameter-expanded Gibbs update of $\kappa$ along the lines of \citet{escobar1995bayesian}.
\item Given the current grouping of $\{\phi_j, \psi_j, \bfpi_j\}_{j = 1}^{n_\ab}$, update the shared parameters $(\phi^*_c, \psi^*_c, \bfpi^*_c)$ of each cluster $c$. Of these, $\bfpi^*_c$ is updated by a Gibbs step by utilizing the multinomial-Dirichlet conjugacy, and, $(\phi^*_c, \psi^*_c)$ is updated by a combination of an independent proposal Metropolis-Hastings update for $\psi^*_c$, followed by a draw of $\phi^*_c$ from  a normal distribution.
\item Save current parameter values as the $s$-th sample draw.
\eenum
}
}
\caption{Schematic description of the Markov chain sampler}
\label{algo1}
\end{algorithm}

\relax

\subsubsection{Estimating $\lambda_\a$ and $\lambda_\b$}
\label{s:first}
One could use any existing aggregation and smoothing technique to estimate the average firing rate curves $\lambda_\a$ and $\lambda_\b$ from A and B trial data. Popular techniques include kernel and spline smoothing as well as more advanced nonparametric methods \citep{kass2003statistical, rao2011gaussian}. However, when either or both of $n_\a$ and $n_\b$ are small, it is important to account for the uncertainty in estimating these curves in the second stage analysis AB trial data. For a full Bayesian analysis, suppose these two unknown curves were assigned prior measures $\Pi_\a$ and $\Pi_\b$ respectively. Then posterior computation can proceed by first updating these priors to posteriors $\Pi_\a(\lambda_\a|\bfX^\a)$ and $\Pi_\b(\lambda_\b|\bfX^\b)$ by using data from only, respectively, the A and the B trials, and, then using these posteriors as new priors for $\lambda_\a$ and $\lambda_\b$ in the second stage analysis of $\bfX^\ab$ detailed above. 

From a practicality perspective, it is most convenient to have the second-stage priors for $\lambda_\a$ and $\lambda_\b$ in the following form:
\beq
\label{gam}
\Pi_e(\lambda_e(t^*_1), \ldots,\lambda_e(t^*_M) | \bfX^e) = \prod_{j = 1}^M \gam(\lambda_e(t^*_m) | a^e_m, b^e_m),~~e \in \{\a, \b\}
\eeq
for some $a^e_m$, $b^e_m$, $m = 1, \ldots, M$, which depend only on $\bfX^e$, i.e., data from the condition $e \in \{\a, \b\}$. Such a structure allows us to fully exploit the conjugacy between the Poisson and the gamma families of distributions. One only needs to extend the MCMC updates detailed in Section \ref{s:mcmc} by making an additional set of draws of $\lambda_e(t^*_m) \sim \gam(a^e_m + \sum_j Z^e_{jm}, b^e_m + n_\ab)$, independently across $e \in \{\a,\b\}$ and $m = 1, \ldots, M$. These draws could be made right after Step 1 of \ref{s:mcmc}.

We fix the parameters $a^e_m$, $b^e_m$ by first smoothing the bin counts of the corresponding single stimulus spike trains. Each spike train is smoothed by using Friedman's super smoother \citep{friedman1984variable}. The average and the variance of the smoothed spike trains are then taken to give the bin specific prior mean ($a^e_m/b^e_m$) and variance ($a^e_m/(b^e_m)^2$) for the second stage analysis.

\begin{rem}
The product nature of the second stage prior in \eqref{gam} is at best a working hypothesis. It may appear less than satisfactory because it introduces additional random variation across bins, even when prior mean and variances are smoothed. One could overcome this deficiency by using importance sampling correction. Suppose $\Pi^*_e(\lambda_e(t^*_1), \ldots, \lambda_e(t^*_M) | \bfX^e)$, $e \in \{a, \b\}$ were the actual prior distributions one had intended to use for the second stage, but the MCMC was run with the product prior given in \eqref{gam} with $a^e_m, b^e_m$ properly chosen so as to match the first two moments under $\Pi^*_e$. One could then obtain Monte Carlo estimates under the intended prior by simply using weighted averages of the saved MCMC draws with the weights being given by the ratio of $\Pi^*_e$ to $\Pi_e$ evaluated at the drawn values of $(\lambda_e(t^*_1), \ldots, \lambda_e(t^*_M))$. 
\end{rem}

\subsection{Prediction}
Inference on $\bbQ$ is best quantified and visualized through the weight curves $\alpha^*$ it is likely to produce in future AB trials. Such $\alpha^*$ may be simulated by making draws from the posterior predictive distribution 
\beq
p(\alpha^* | \bfX^\ab, \bfX^\a, \bfX^\b) = \int p(\alpha^*|\bftheta) p(\bftheta | \bfX^\ab,\bfX^\a, \bfX^b)d\bftheta
\label{pred}
\eeq
where $\bftheta = (\kappa, \{\bfeta_j, \phi_j, \psi_j, \bfpi_j, \ell_j\}_{j = 1}^{n_\ab})$ denotes the ensemble of all model parameters that are included in the MCMC sampling of Section \ref{s:mcmc}. Draws of $\bfeta^*$ from \eqref{pred} may be made by drawing one $\alpha^*$ from $p(\alpha^* | \bftheta)$ for each saved draw of $\bftheta$ from the Markov chain sampler. Let $\phi^*$, $\psi^*$, $\bfpi^*$, $\ell^*$ and $\bfeta^*$ denote the latent quantities associated with $\alpha^*$ as in \eqref{pq1}-\eqref{pq3}. Notice that,
\begin{align}
p(\alpha^* | \bftheta)  = p(\alpha^* | \bfeta^*, \phi^*, \psi^*, \ell^*) & p (\bfeta^* | \phi^*, \psi^*, \ell^*)p(\ell^*|\bfpi^*) \label{pred1}\\
& \times 
p(\phi^*, \psi^*, \bfpi^* | \{\phi_j, \psi_j, \bfpi_j\}_{j = 1}^{n_\ab})\label{pred2}
\end{align}
and hence a draw of $\alpha^*$ from $p(\alpha^*|\theta)$ can be made by making draws from the four conditional distributions on the right hand side, proceeding sequentially from right to left. It is easy to make draws from the three posterior distributions appearing on \eqref{pred1} as they are governed purely by the relationships in \eqref{pq1}-\eqref{pq3}. The conditional distribution in \eqref{pred2}, again by the P\'olya urn scheme representation of the Dirichlet process, is given by:
\[
p(\phi^*, \psi^*, \bfpi^* | \{\phi_j, \psi_j, \bfpi_j\}_{j = 1}^{n_\ab}) = \frac{\kappa}{\kappa + n_\ab} G_\kappa + \frac{1}{\kappa + n_\ab} \sum_{c = 1}^K \delta_{(\phi^*_c, \psi^*_c, \bfpi^*_c)},
\]
where $K$ denote the number of distinct elements $(\phi^*_c, \psi^*_c, \bfpi^*_c)$, $c = 1, \ldots, K$, among the collection $\{(\phi_j, \psi_j, \bfpi_j): j = 1, \ldots, n_\ab\}$.

\section{Numerical studies}
\subsection{Case studies with synthetic data}
\label{case studies}

We report here three simulation experiments in which we assessed the scope of the DAPP model and the accuracy of the statistical method introduced in this work. Each experiment consisted of one cell with a distinct form of second order stochasticity:
\begin{description}
\item[Experiment 1 (random selection).] The cell always produces flat weight curves $\alpha(t) \equiv \alpha$, with the magnitude $\alpha$ drawn either uniformly from $(0.05, 0.25)$ with probability 60\% or uniformly from $(0.85, 0.95)$ with probability 40\%.
\item[Experiment 2 (dynamic averaging with random period).] The cell always produces sinusoidal weight curves $\alpha(t) = 0.01 + 0.49 \{1 + \sin(2\pi\frac{a + t}{b})\}$ which oscillate between 0.01 and 0.99, where the random period $b$ (in ms) is drawn uniformly from the range (400, 1000) and the random shift (also in ms) is drawn uniformly from $(0, b)$. 
\item[Experiment 3 (mixture of flat averaging and dynamic averaging).] The cell produces a 50-50 mixture of flat and sinusoidal weight curves. For the flat curves, the time invariant magnitude is drawn uniformly from (0.4, 0.7). The sinusoidal curves are again of the form $\alpha(t) = 0.01 + 0.49 \{1 + \sin(2\pi\frac{a + t}{b})\}$, but with the random period now drawn uniformly from the range (320, 340) and the random shift is drawn uniformly from (0, $b$).
\end{description}

\begin{figure}[!t]
\centering
\graphsc{0.15}{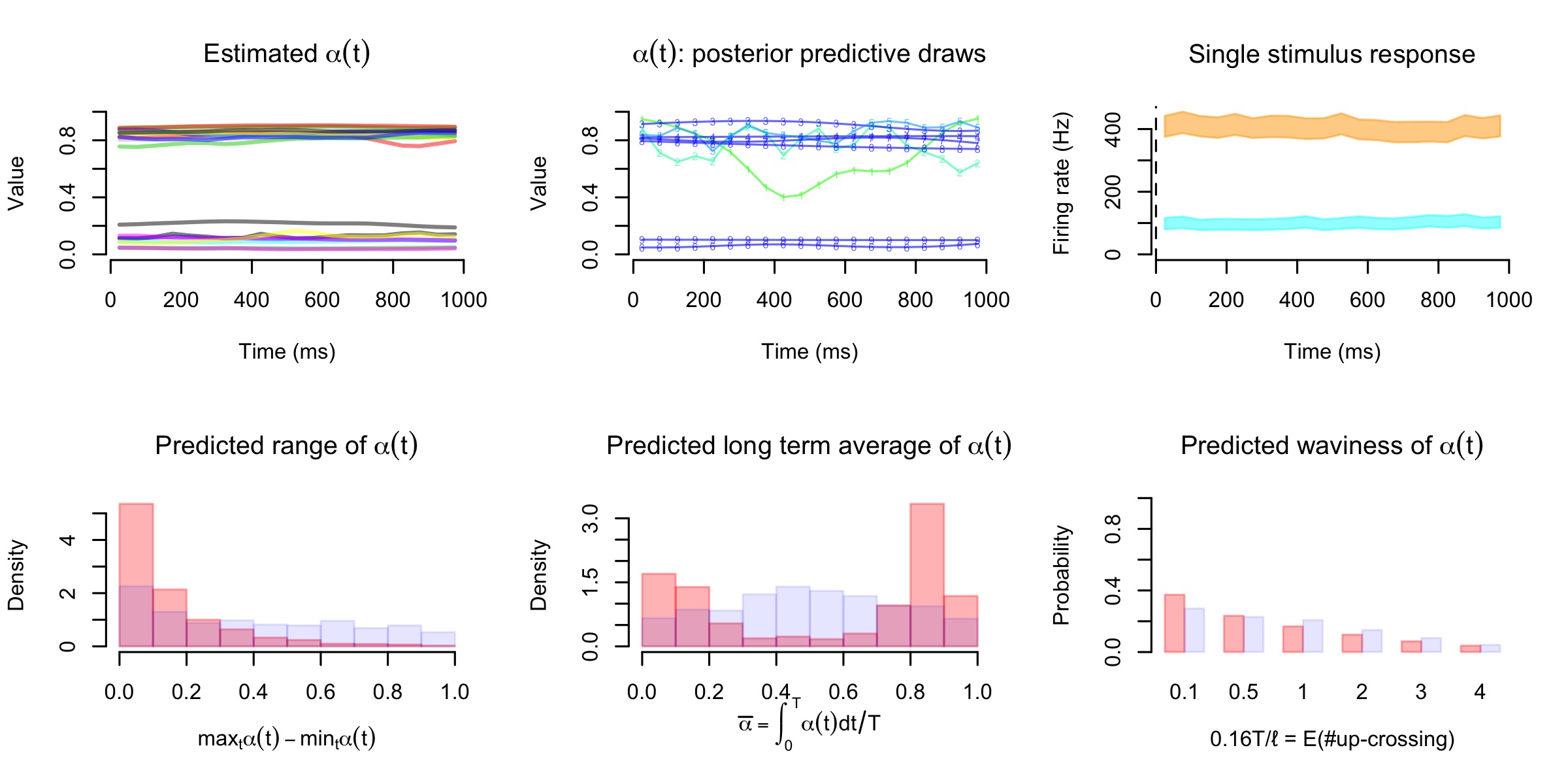}
\includegraphics[trim={0 22.5cm 0 0},clip,height=3cm]{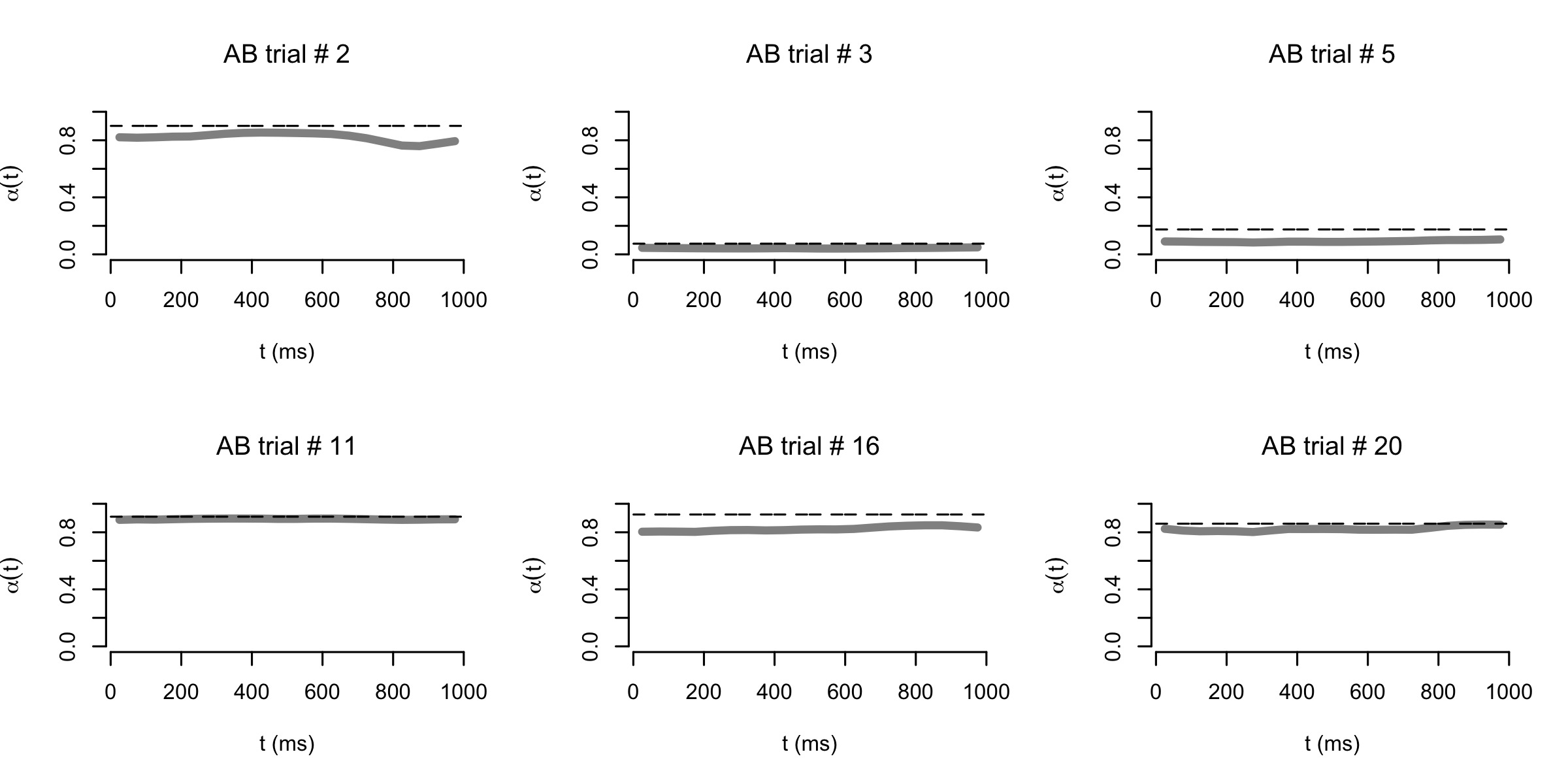}
\caption{Analysis of Experiment 1 data consisting of 20 flat $\alpha$ curves. Eight of them had constant magnitudes in the range (0.05, 0.25) and the other 12 in the range (0.85, 0.95). Top row displays three posterior summaries: 1) estimates of the 20 $\alpha$ curves based on observed data; 2) 9 posterior predictive draws of $\alpha$; and, 3) estimates and 95\% credible intervals for $\lambda_\a(t)$ and $\lambda_\b(t)$. Second row shows (in pale red) the posterior predictive distributions of 1) $\mathrm{range}(\alpha) = \max_{t \in [0,T]} \alpha(t) - \min_{t \in [0,T]} \alpha(t)$; 2) the long term average $\bar \alpha$; and, 3) the waviness as captured by the expected up-crossing count. The histograms shown are created from 1000 posterior predictive draws of $\alpha$ and the underlying quantities. The pale blue histograms show the same but with draws of $\alpha$ made from the prior distribution. The third row compares the estimated $\alpha$ (solid line) against the true curve (dashed line) for 3 randomly chosen AB trials.} 
\label{synth 1}
\efig

For each experiment, we assumed $\lambda_\a \equiv 400$ (in Hz) and $\lambda_\b= 100$ and simulated 20 A trials, 20 B trials and 20 AB trials with a common response horizon of $T = 1000$ (in ms). The resulting 60 spike trains were analyzed under the DAPP model with a 50 ms bin-width used for time discretization. To assess what the DAPP model learned about the nature of the second order stochasticity, we focus on the posterior predictive summaries of three broad features of the weight curves: 1) the range of $\alpha$ defined as $\mathrm{range}(\alpha) = \max_{t \in [0,T]} \alpha(t) - \min_{t \in [0,T]} \alpha(t)$; 2) the long term average $\bar \alpha$; and, 3) the waviness as captured by the expected up-crossing count $0.16 T / \ell$, with $\ell$ denoting the characteristic length-scale underlying $\alpha$. 
Figures \ref{synth 1}, \ref{synth 2} and \ref{synth 3} show the results of our data analysis for these three synthetic cases.

\begin{figure}[!tp]
\centering
\graphsc{0.15}{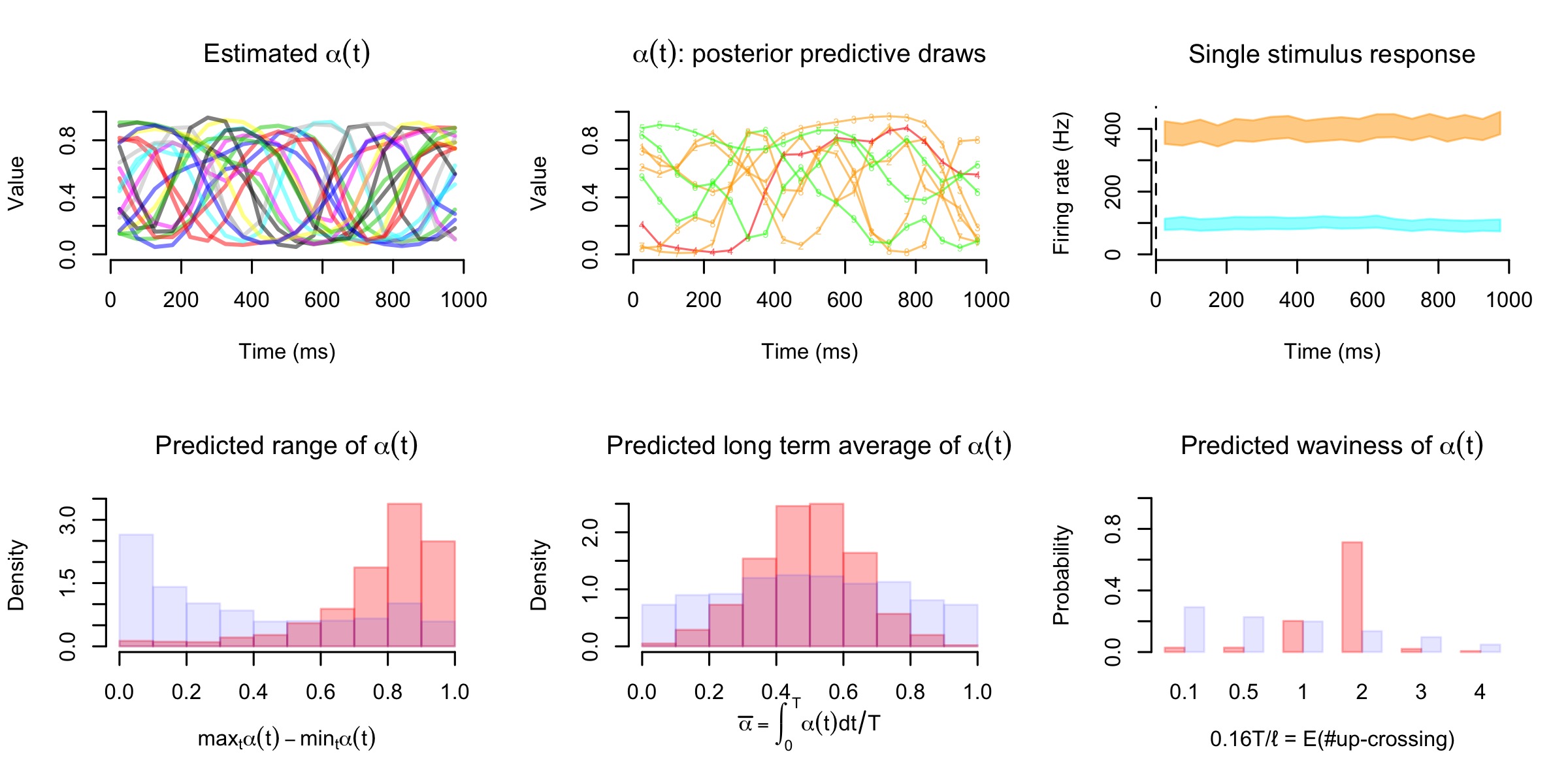}
\includegraphics[trim={0 22.5cm 0 0},clip,height=3cm]{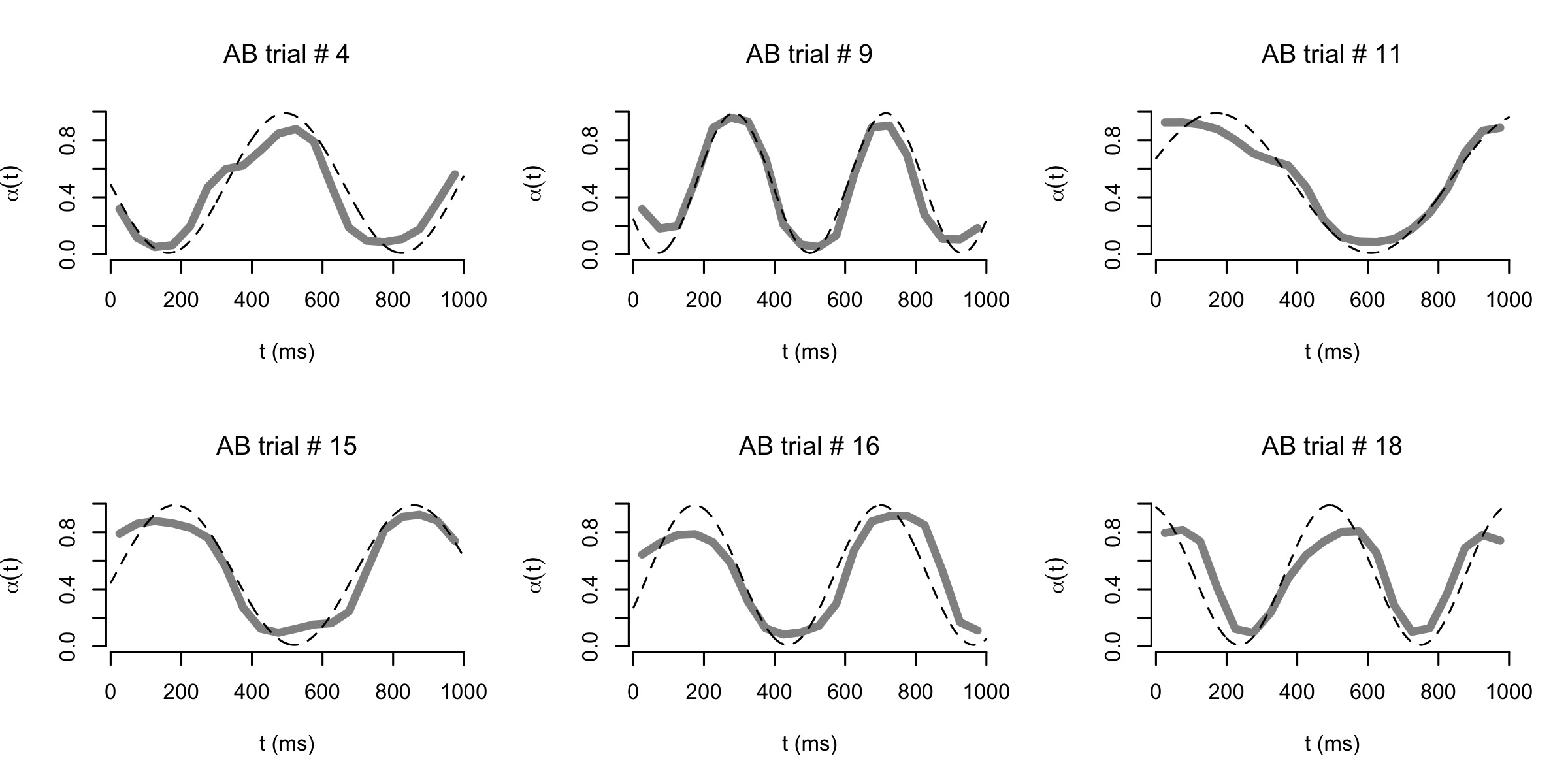}
\caption{Analysis of Experiment 2 data consisting of 20 sinusoidal $\alpha$ curves, each swinging between 0.01 and 0.99 with a period (in ms) drawn randomly from the interval (400, 1000).} 
\label{synth 2}
\efig

\begin{figure}[!tp]
\centering
\graphsc{0.15}{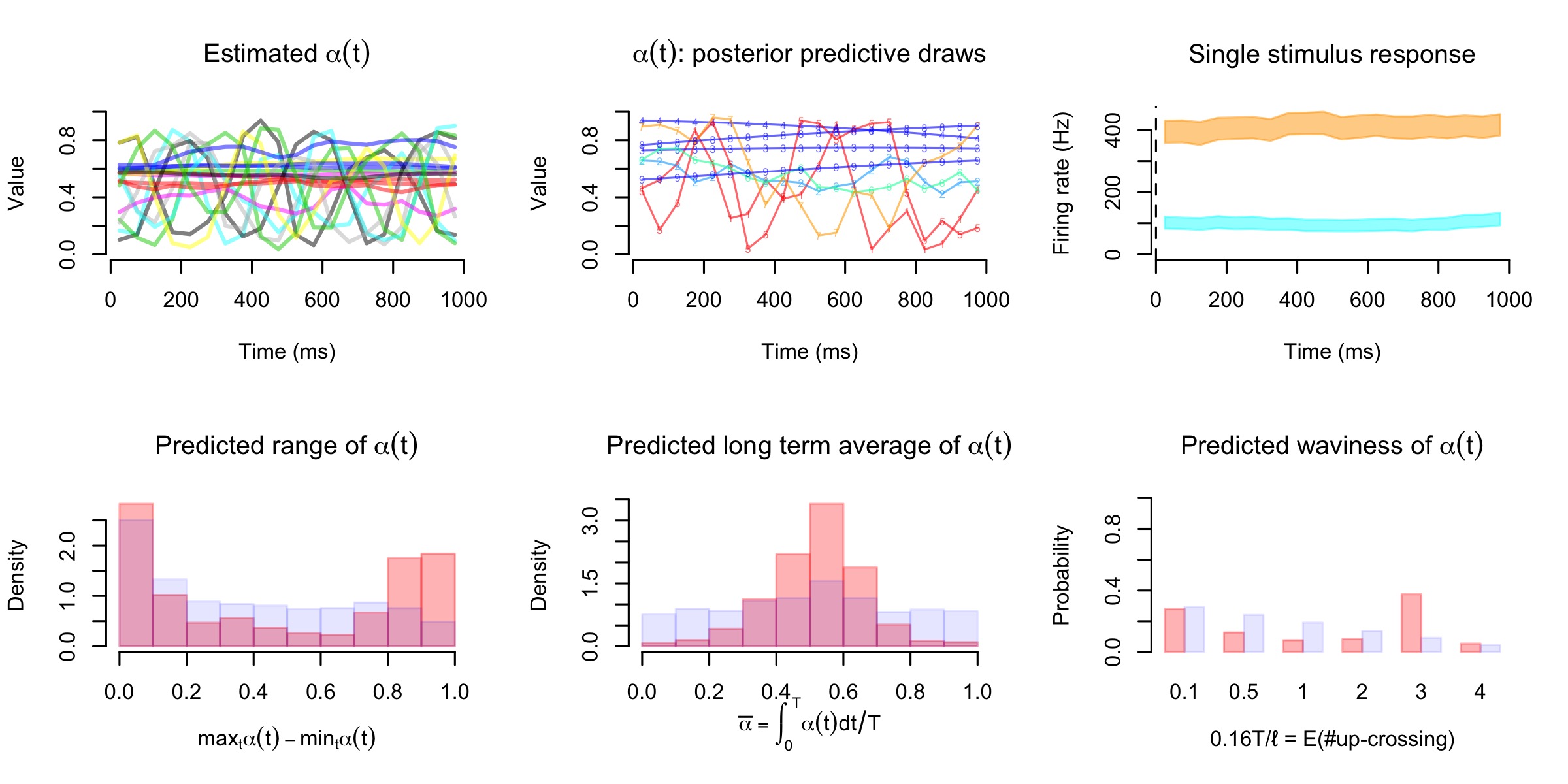}
\includegraphics[trim={0 22.5cm 0 0},clip,height=3cm]{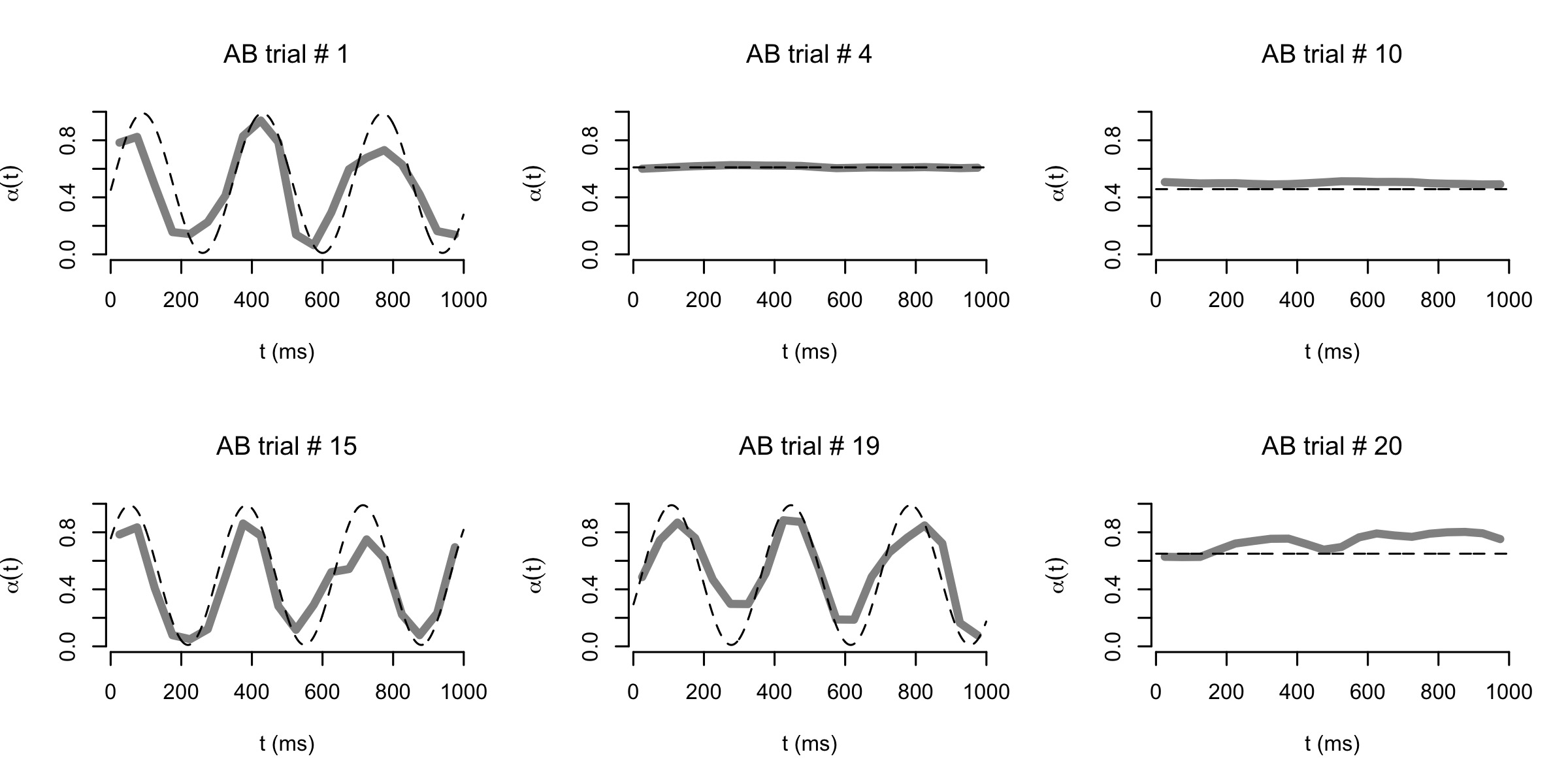}
\caption{Analysis of Experiment 3 data consisting of 11 flat (magnitude between 0.4 and 0.7) and 9 sinusoidal $\alpha$ curves (swing between (0.01, 0.99), period between (320, 340) ms).} 
\label{synth 3}
\efig

From the figures we conclude that the DAPP model is able to correctly recover the broad features of the second order stochasticity in each case. This is particularly evident from the second row plots. For Experiment 1, the posterior predictive histogram of range$(\alpha$) is peaked near zero and the posterior predictive distribution of the expected up-crossing count is also peaked at the smallest allowable value of 0.1. Both pictures suggest the model correctly learned that cell 1 produces mostly flat curves. Furthermore, the posterior predictive histogram of  $\bar \alpha$ is bimodal with peaks near zero and one, indicating that the model correctly learned nearly half of the flat weight curves are close to zero while the other half are close to one. 

For Experiment 2, the posterior predictive histogram of range($\alpha$) is peaked near one and the distribution of the expected up-crossing count is peaked at 1 and 2, suggesting, quite accurately, that the cell mostly produces wavy weight curves that oscillate from one extreme to the other with a period that is mostly in the range of $T/2 = 500$ to $T/1 = 1000$ ms. 

For Experiment 3, the posterior predictive histogram of range($\alpha$) is bimodal with peaks near zero and one, and, the posterior predictive distribution of the expected up-crossing count is also bimodal with peaks at 0.1 and 3. These collectively suggest, again quite accurately, that the cell produces a mix of flat and wavy curves, where the latter ones oscillate the entire range with a period of about $T/3 = 333$ ms.

\subsection{Second order stochasticity of inferior colliculus neurons}
\label{s:real}
We report here results of the DAPP model analysis of the spiking activity of the three inferior colliculus (IC) cells referred to in Figures \ref{fig 3 cells} and \ref{reals-total}. We focus on these three speciﬁc cells to assess whether the DAPP model picks up different modes of naturally occurring second order stochasticity. 

The neural data reported here comes from the data set described in \citet{caruso2018single}.  Briefly, the activity of individual neurons in the inferior colliculus was recorded while two monkeys listened for and made eye movements to the locations of sounds.  Each trial began with the onset of a visual target located straight ahead, which the monkey was required to fixate on before the trial could proceed.  Then, either one or two sounds were presented.  These sounds stayed on for 600-1000 ms, before the fixation light was extinguished, cuing the monkey to make eye movements to each sound (one if one sound, two if two sounds).  The dual sounds were located at either (-24 deg, +6 deg) or (-6 deg, +24 deg) horizontally, and consisted of bandpass noise with different center frequencies (742 Hz and another frequency that differed by a ratio of 1.22 or an integer power of that ratio.  Single sounds were drawn from the same set of locations and frequencies that were used on the dual sound trials.  The neural activity was analyzed during the first 600-1000 ms of the epoch that the sounds were on but the monkey was maintaining fixation. All conditions were randomly interleaved.

Of the three cells shown here, Cell 1 comes from monkey P and involved a 903 Hz sound (A) at 24 degrees to the left and 742 Hz sound (B) at 6 degrees to the right.  There were 29 A trials, 21 B trials and 34 AB trials, and the analysis was conducted on a 600 ms response period. Cell 2 and Cell 3 recordings were made from monkey Y.  For Cell 2, sound B was centered at 742 Hz and located at 6 degrees to the left and sound A was 609 Hz located at 24 degrees to the right. Cell 3 involved the same sound B frequency and location (742 Hz, 6 degrees left), but sound A was a 500 Hz sound (24 degrees right). Cell 2's response period was 600 ms whereas cell 3 had a response period of length 1000 ms. Cell 2 had 7 trials each in conditions A, B and AB. Cell 3 had 15 A trials, 11 B trials and 37 AB trials.

An underlying assumption of the DAPP model is that the AB firing rates lie within the range defined by the A and B firing rates. For the IC cells, we assess the validity of this assumption through whole-trial spike counts. Recall that Figure \ref{reals-total} shows smoothed histograms of whole trial spike counts grouped by conditions A, B and AB. Notice that for each cell, the AB distribution appears to sit between the distributions under conditions A and B, conforming with the DAPP assumption. 

\begin{figure}[!tp]
\centering
\subfloat[Cell 1]{\includegraphics[height = 6cm, width = 12cm]{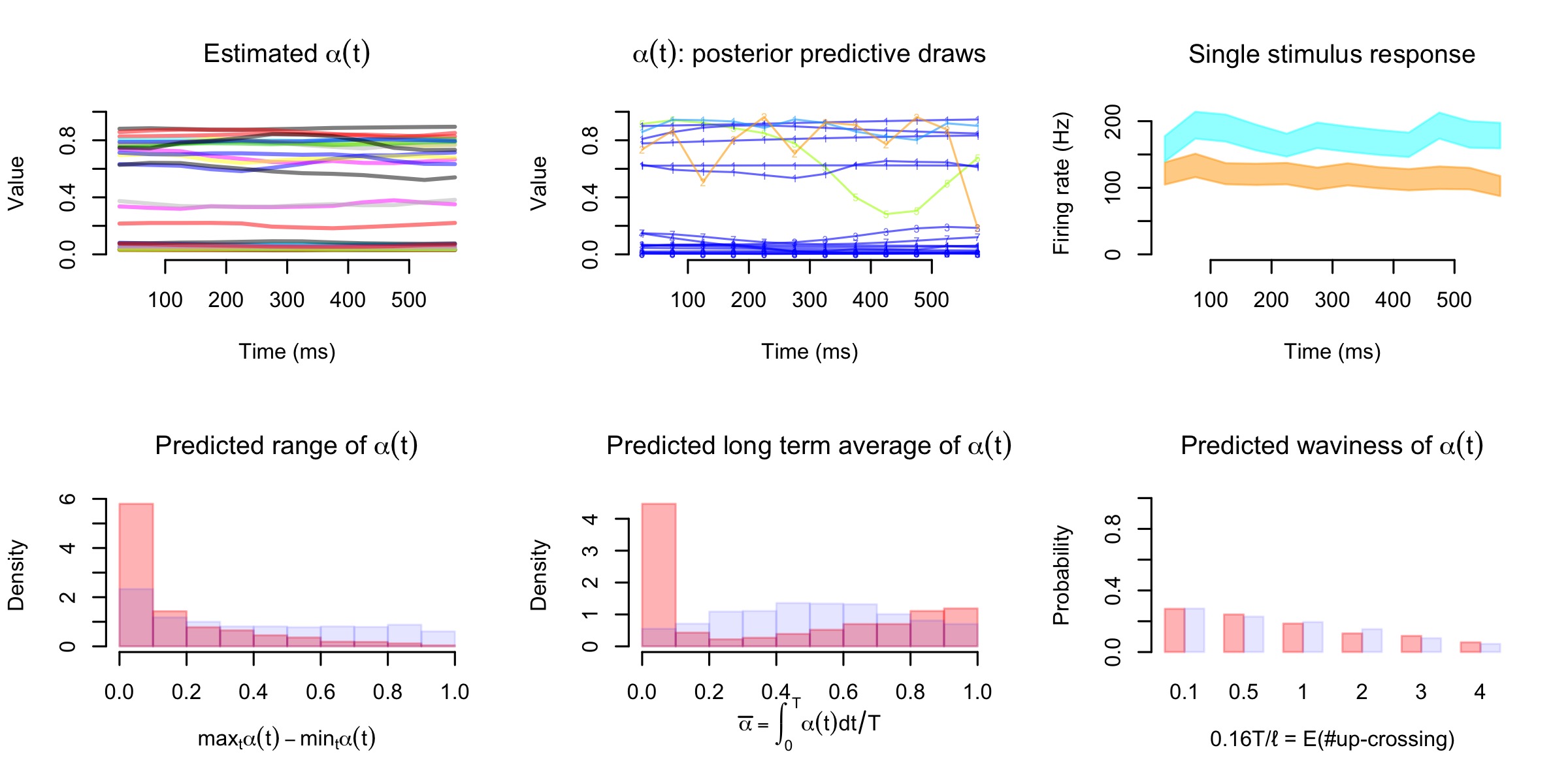}}\\
\subfloat[Cell 2]{\includegraphics[height = 6cm, width = 12cm]{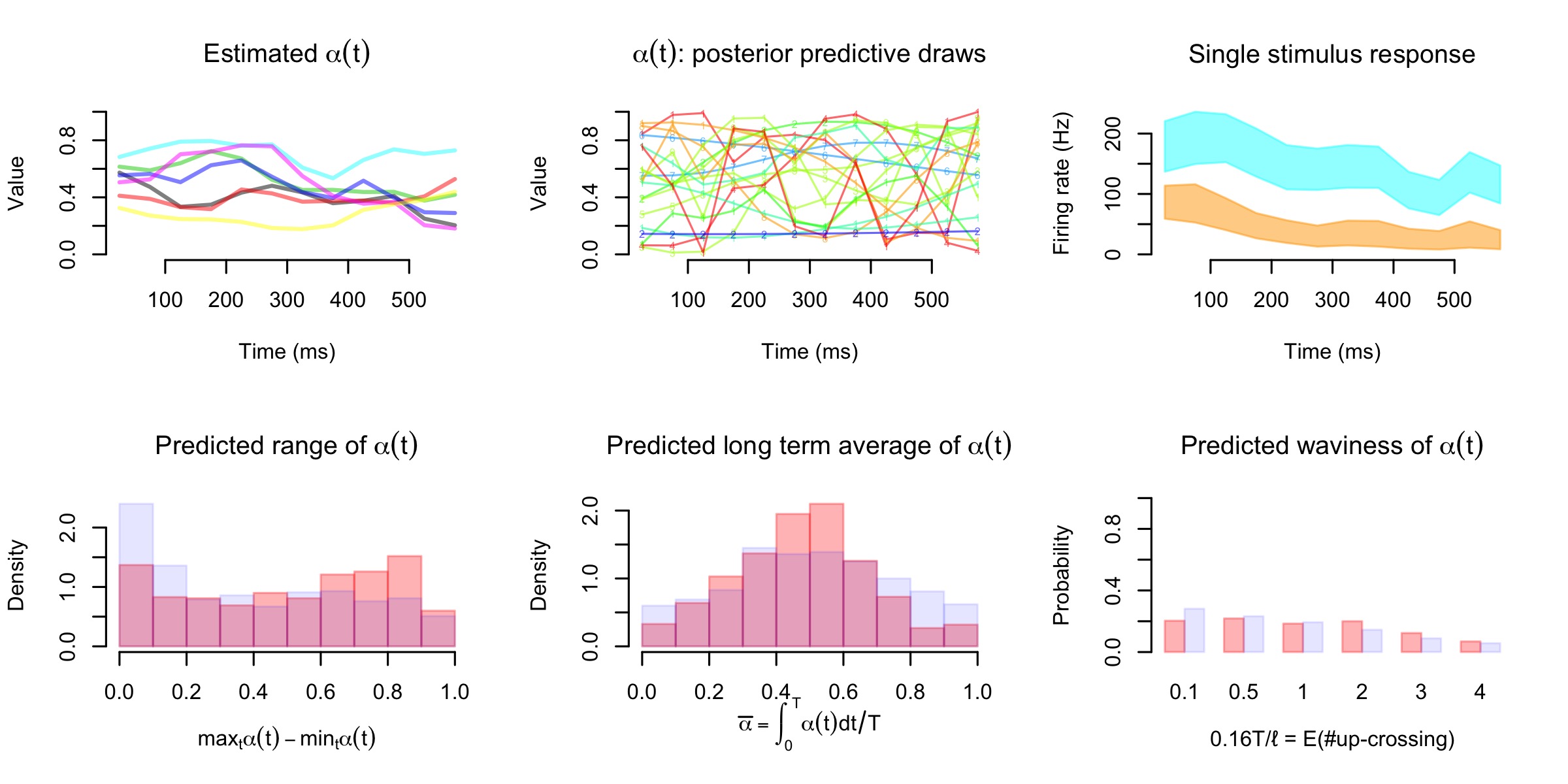}}\\
\subfloat[Cell 3]{\includegraphics[height = 6cm, width = 12cm]{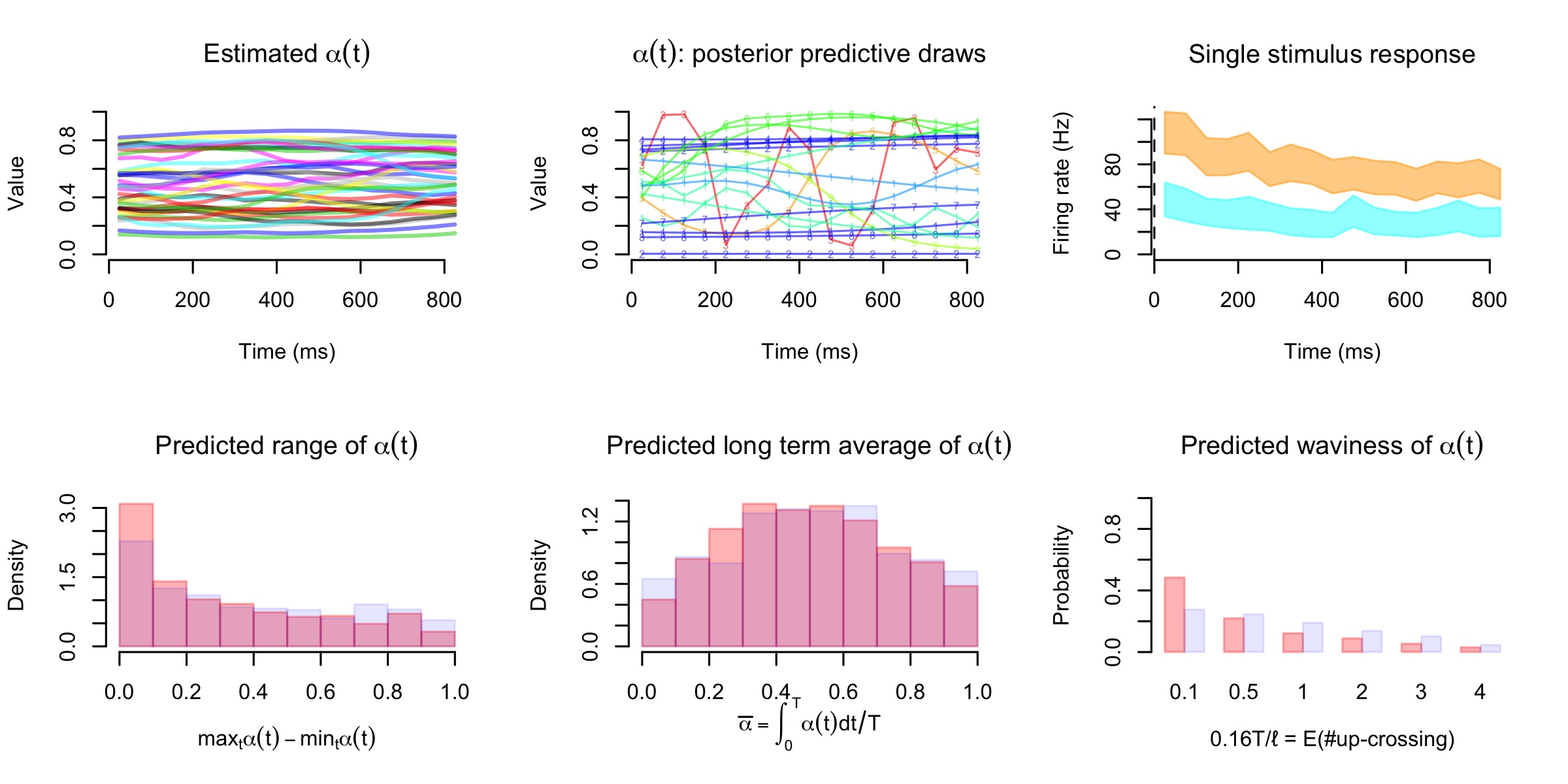}}
\caption{Posterior inference for three example IC cells.} 
\label{reals-dapp}
\efig

Figure \ref{reals-dapp} visualizes the DAPP analysis results for the 3 IC cells. It is immediately apparent that the three cells exhibit different patterns of second order stochasticity. Cell 1 exhibits a second order stochasticity pattern similar to random selection. It has nearly 50\% chance of producing an $\alpha$ curve that is flat and very close to zero, whereas, with nearly 30\% probability it would produce a flat $\alpha$ curve with magnitude in the range (0.6, 1). That is, when exposed to both A and B sounds, almost half of the time the cell would respond like it is responding only to sound B. But in about every third trial its response will be more resembling of its sound A spiking activity, although this latter resemblance is less exact.

In contrast, cell 2 appears to have a higher likelihood of producing a wavy $\alpha$ curve. The posterior predictive distribution of range($\alpha$) is bimodal with 40\% mass concentrated in the interval (0.6, 0.9). Recall that range($\alpha) \ge 0.6$ means that the weight $\alpha(t)$ attached to $\lambda_\a(t)$ dynamically swings from 0-20\% to 80-100\% (or the reverse) -- a phenomenon consistent with within trial random interleaving. However, the posterior distribution of expected up-crossing count does not support many oscillations. The likely scenario is that a future AB trial firing rate curve would exhibit one swing from being nearly A like to nearly B like. Notice that the there is an overall lack of high concentration of these posterior distributions underlining that the DAPP model parameters were not learned with high confidence. This is not surprising given that the cell had only 7 trials in each condition. 

Cell 3 appears to have a high propensity of producing flat $\alpha$ curves (50\% chance of no up-crossing) with the long term average distributed over the entire interval $(0,1)$ with a mild concentration near the center. This pattern is entirely different from the concept of random selection or interleaving. Here, on any AB trial, the cell appears to exhibit a spiking activity with a firing rate that is a non-dynamic weighted average of $\lambda_\a$ and $\lambda_\b$, but the weights assigned to the two pure sound average firing rates vary across trials with a concentration around the case where both sounds are given equal weights.

\section{Discussion}
\label{s:disc}

In this paper we have introduced a novel concept of second order stochasticity in neuronal firing rates in response to a stimuli bundle. The very definition of second order stochasticity, rooted in the information-preserving, stochastic variation of the firing rate curve from one trial to the next, rules out the commonly used time-and-trial aggregated statistical methods for analyzing spike train data. We have developed a detailed point pattern model, namely the DAPP model, based on the assumption of stochastically varying, dynamic averaging of single stimulus firing rate curves. Our model is generative in nature. The fitted model can be used to draw inference on how the cell is likely to respond in future hypothetical trials under the stimuli bundle exposure.

Our treatment of second order stochasticity leaves room for many further developments. The overarching assumption of dynamic averaging can only explain special kinds of second order stochasticity where under the stimuli bundle exposure, the overall firing rate of the cell resides in between the rates it exhibits under each individual stimulus. Stimuli bundles that evoke either enhancement or suppression of activity, i.e., producing rates outside the range of single stimulus response rates, cannot be analyzed with the DAPP model.

Our model assumes spike counts are Poisson distributed with possibly time varying firing rate curves. It is known that the Poisson assumption does not always provide the best fit to inter-spiking interval distributions observed in reality. One issue with the Poisson assumption is its inability to account for the refractory period which is a short time gap immediately after a spike during which the neuron cannot fire again no matter what stimulus is presented to it. However, this is not a big issue in our applications where spiking activity is aggregated in 50 ms time bins, which is much longer time scale than the typical length of a refractory period which is usually no more than 2 ms. A second issue with the Poisson assumption is its inability to account for {\it over-dispersion} where the variance of the spiking activity is larger than its mean.  This may be accounted for by extending our DAPP model where the Poisson assumption is replaced with a negative-binomial assumption. However, defining and analyzing negative-binomial point processes with smooth rate curves pose serious technical challenges that are beyond the scope of the current paper.

In applying the DAPP model and method developed here, one needs to choose the binning interval width to carry out the time discretization of spiking activity. While shorter bins allow more flexible estimation of the time varying dynamics of the $\alpha$ curves, the increased number of bins adds to computing cost. Shorter bins, with fewer spike counts in each, also pose some difficulty to our two-stage estimation process for the $\lambda_\a$ and $\lambda_\b$ curves. Specifically, the second stage prior in \eqref{gam} which assumes conditional independence of the curve values across bins, do not offer adequate smoothing. There is potential to remedy this problem by replacing the product gamma prior in \eqref{gam} with an autoregressive gamma prior \citep{wolpert1998poisson}. Our choice of 50 ms for bin width was based on our understanding of the scale of IC firing rate. We also repeated the analyses in Section \ref{s:real} with 25 ms bins and the results were robust to this change in bin width.

These challenges notwithstanding, the DAPP analysis framework presented in this paper offers an important first step toward understanding, modeling and estimating second order stochasticity. \citet{caruso2018single} give strong evidence of the prevalence of second order stochasticity in the primate brain, as well as, of the utility of the DAPP analysis in cataloging various modes of such stochastic variation. Clearly, it will take a system level understanding of neural computing to completely describe how the brain might represent multiple simultaneous signals. The cell level DAPP analysis promises to be an important building block toward such a goal.

\begin{appendices}
\section{Details of Markov chain sampling}
\label{app-mcmc}

\paragraph{Step 1. Gibbs update of $(\bfZ^\a, \bfZ^\b, \bfY^\a, \bfY^\b )$.} Since
\begin{align*}
p(\bfZ^\a, \bfZ^\b, & \bfY^\a, \bfY^\b~|~\kappa, \{\bfeta_j, \phi_j, \psi_j, \bfpi_j, \ell_j\}_{j = 1}^{n_\ab}, \bfX^\ab, \lambda_\a, \lambda_\b)\\
& = p(\bfY^\a, \bfY^\b ~|~ \bfX^\ab, \{\bfeta_j\}_{j = 1}^{n_\ab}, \lambda_\a, \lambda_\b) \\
&~~~~~~~~\times p(\bfZ^\a | \bfY^\a, \{\bfeta_j\}_{j = 1}^{n_\ab}, \lambda_\a) \times p(\bfZ^\b | \bfY^\b, \{\bfeta_j\}_{j = 1}^{n_\ab}, \lambda_\b),
\end{align*}
one can draw $(\bfZ^\a, \bfZ^\b, \bfY^\a, \bfY^\b)$ from their joint conditional posterior by drawing, in succession, from the three conditional distributions on the right of the above display. This proceeds in three steps for each $j = 1, \ldots, n_\ab$, $m  = 1, \ldots, M$:
\benum
\item Draw $Y^\a_{jm} \sim \bin(X^\ab_{jm}, \frac{\alpha_{jm} \lambda_\a(t^*_m)}{\alpha_{jm} \lambda_\a(t^*_m) + (1-\alpha_{jm}) \lambda_\b(t^*_m)})$ and set $Y^\b_{jm} = X^\ab_{jm} - Y^\a_{jm}$, where $\alpha_{jm} = 1/(1 + e^{-\eta_{jm}})$.
\item Draw $\bar Z^\a_{jm} \sim \pois(w(1 - \alpha_j(t^*_m))\lambda_\a(t^*_m))$ and set $Z^\a_{jm} = Y^\a_{jm} + \bar Z^\a_{jm}$.
\item Draw $\bar Z^\b_{jm} \sim \pois(w\alpha_j(t^*_m)\lambda_\b(t^*_m))$ and set $Z^\b_{jm} = Y^\b_{jm} + \bar Z^\b_{jm}$
\eenum

\paragraph{Step 2. Parameter-expanded Gibbs update of $\{\bfeta_j, \ell_j\}_{j = 1}^{n_\ab}$.} We can write
\begin{align*}
p(&\{\bfeta_j, \ell_j\}_{j = 1}^{n_\ab} ~|~\bfZ^\a, \bfZ^\b, \bfY^\a, \bfY^\b, \kappa, \{\phi_j, \psi_j, \bfpi_j\}_{j = 1}^{n_\ab}, \bfX^\ab, \lambda_\a, \lambda_\b)\\
& ~~~~=p(\{\bfeta_j, \ell_j\}_{j = 1}^{n_\ab} ~|~\bfZ^\a, \bfZ^\b, \bfY^\a, \bfY^\b, \{\phi_j, \psi_j,\bfpi_j\}_{j = 1}^{n_\ab})\\
& ~~~~\propto \prod_{j = 1}^{n_\ab} \left \{\prod_{m = 1}^M \bin\left(y^*_{jm} \bigg| N_{jm}, \frac{1}{1 + e^{-\eta_{jm}}}\right) \right\} \nm(\bfeta_j | \phi_j{\bf1}, \psi_j \bfC_{\ell_j}) P_{\scL, \bfpi_j}(\ell_j),
\end{align*}
where $y^*_{jm} = Y^\a_{jm} + Z^\b_{jm} - Y^\b_{jm}$, $N_{jm} = Z^\a_{jm} + Z^\b_{jm}$, ${\bf 1}$ is the $M$-dimensional vector of ones, and, $\bfC_{\ell_j}$ is an $M\times M$ covariance matrix with $(m,m')$-th element $C^\se_{\ell_j}(t^*_m, t^*_{m'})$. The factorization over $j$ allows us to make parallel Gibbs updates of $(\bfeta_j, \ell_j)$ across $j = 1, \ldots, n_{\ab}$, which can be accomplished by the P\'olya-gamma parameter augmentation trick of \citet{polson2013bayesian}. This proceeds in two steps:
\benum
\item Draw $\omega_{jm} \sim \textit{PG}(N_{jm}, \eta_{jm})$, independently across $m = 1, \ldots, M$. Here $\textit{PG}(b, c)$ denote the P\'olya-gamma distribution with shape parameter $b$ and tilting parameter $c$. Set ${\bf\Omega}_j = \diag(\omega_{j1}, \ldots, \omega_{jM})$.
\item Update $(\bfeta_j, \ell_j)$ based on the local model:
\[
\bar \bfy_j | (\bfeta_j, \ell_j) \sim \nm(\bfeta_{j}, {\bf\Omega}_{j}^{-1}),~~~~\bfeta_j | \ell_j \sim \nm(\phi_j{\bf 1}, \psi_j \bfC_{\ell_j}),~~~~\ell_j \sim P_{\scL, \bfpi_j}
\]
where $\bar \bfy_j = (\bar y_{j1}, \ldots, \bar y_{jM})$ with $\bar y_{jm} = y^*_{jm} - N_{jm}/2$. That is, with $\bfpi_j = (\pi_{j1}, \ldots, \pi_{jL})$, one first draws $\ell_j$ from $\{\ell^*_1, \ldots, \ell^*_L\}$ according to probabilities $(q^j_1, \ldots, q^j_L)$ where $q^j_i \propto \pi_{ji} \nm(\bar y_j | \phi_j {\bf 1}, \psi_j \bfC_{\ell^*_i} + {\bf\Omega}_j^{-1})$. Then one draws $\bfeta_j$ from $\nm(\bf m, \bf S)$ where $\bfS = ({\bf\Omega}_j + \psi_j^{-1} \bfC_{\ell_j}^{-1})^{-1}$ and $\bfm = \bfS({\bf\Omega}_j \bar \bfy_j + \phi_j \psi_j^{-1} \bfC_{\ell_j}^{-1}{\bf1})$
\eenum

\paragraph{Step 3. Parameter-expanded Gibbs update of $\{\phi_j, \psi_j, \bfpi_j\}_{j = 1}^{n_\ab}$.} We sequence through $j = 1, \ldots, n_\ab$ and update $(\psi_j, \psi_j)$ given all other parameter values. At any such instance $j = i \in \{1, \ldots, n_\ab\}$,
%
\begin{align*}
p(\phi_i, & \psi_i, \bfpi_i | \bfZ^\a, \bfZ^\b, \bfY^\a, \bfY^\b, \kappa, \{\bfeta_j, \phi_j, \psi_j, \bfpi_j ,\ell_j\}_{j \ne i}, \bfeta_i, \ell_i, \bfX^\ab, \lambda_\a, \lambda_\b)\\
& = p(\phi_i, \psi_i, \bfpi_i | \kappa, \{\phi_j, \psi_j, \bfpi_j\}_{j \ne i}, \bfeta_i, \ell_i)\\
& \propto p(\bfeta_i | \phi_i, \psi_i, \ell_i) p(\ell_i | \bfpi_i) p(\phi_i, \psi_i, \bfpi_i | \kappa, \{\phi_j, \psi_j, \bfpi_j\}_{j  \ne i}) 
\end{align*}
with $p(\bfeta_i | \phi_i, \psi_i, \ell_i) = \nm(\bfeta_i | \phi_i {\bf 1}, \psi_i \bfC_{\ell_i})$, and, by the P\'olya urn scheme representation of the Dirichlet process,
\[
p(\phi_i, \psi_i, \bfpi_i | \kappa, \{\phi_j, \psi_j, \bfpi_j\}_{j \ne i}) = \frac{1}{\kappa + n_\ab - 1} \left\{\kappa G_\kappa + \sum_{c = 1}^{K^-} n_{-i,c} \delta_{(\phi^*_c, \psi^*_c, \bfpi^*_c)}\right\}
\]
where $(\phi^*_c, \psi^*_c, \bfpi^*_c)$, $c = 1, \ldots, K^-$, are the distinct elements in the collection $\{(\psi_j, \psi_j, \bfpi_j) : j \ne i\}$ with $n_{-i,c}$ giving the number of times the $c$-th distinct element appears in the collection. Hence, Algorithm 8 of \cite{neal2000markov}, with a given auxiliary sample size $r$, can be used to update $(\phi_i, \psi_i, \bfpi_i)$ as follows:
\benum
\item Draw $r$ additional pairs $(\phi^*_{K^- + h}, \psi^*_{K^- + h}, \bfpi^*_{K^- + h}) \sim G_\kappa$, $h = 1, \ldots, r$. If $(\phi_i, \psi_i, \bfpi_i) \not\in \{(\phi^*_c, \psi^*_c, \bfpi^*_c): 1 \le c \le K^-\}$, then replace the first additional draw $(\phi^*_{K^- + 1}, \psi^*_{K^- + 1}, \bfpi^*_{K^- + 1})$ with $(\phi_i, \psi_i,\bfpi_i)$. 
\item Draw an index $c_i$ from $\{1, \ldots, K^- + r\}$ according to probabilities $p_c \propto s_c \cdot P_{\scL, \bfpi^*_c}(\ell_i) \cdot\nm(\bfeta_i | \phi^*_c {\bf 1}, \psi^*_c \bfC_{\ell_i})$, $1 \le c \le K^- + r$, where $s_c = n_{-i,c}$ for $c = 1, \ldots, K^-$, and, $s_c = \kappa / r$ for $c  = K^- + 1, \ldots, K^- + r$.
\eenum

\paragraph{Step 4. Parameter-expanded Gibbs update of $\kappa$.} Now, let $(\phi^*_c, \psi^*_c, \bfpi^*_c)$, $c = 1, \ldots, K$, denote the distinct elements in the full collection $\{(\phi_j, \psi_j, \bfpi_j): j = 1, \ldots, n_\ab\}$. Following \cite{escobar1995bayesian}, we write
\begin{align*}
p(\kappa & | \bfZ^\a, \bfZ^\b, \bfY^\a, \bfY^\b, \{\bfeta_j, \phi_j, \psi_j, \bfpi_j, \ell_j\}_{j = 1}^{n_\ab}, \bfX^\ab, \lambda_\a, \lambda_\b)\\
& \propto p(\kappa) p(\{\phi_j, \psi_j, \bfpi_j\}_{j = 1}^{n_\ab} | \kappa)\\
& \propto e^{-\kappa} \frac{\kappa^K}{\prod_{j = 1}^{n_\ab} (\kappa + j - 1)} \prod_{c = 1}^K \{\kappa (1 - \psi^*_c)^{\kappa - 1}\}\\
& \propto B(\kappa, n_\ab) \kappa^{2K} e^{-b \kappa}\\
& = \int_0^1 \omega^{\kappa - 1}(1 - \omega)^{n_\ab - 1} \kappa^{2K} e^{-b\kappa} d\omega
\end{align*}
where $b = 1 - \sum_{c = 1}^K \log (1 - \psi^*_c)$ and $B(a, b) = \int_0^1 \omega^{a - 1}(1 - \omega)^{b - 1}d\omega = \Gamma(a)\Gamma(b)/\Gamma(a + b)$ denotes the beta function. Therefore the conditional posterior density function of $\kappa$ is identical to its marginal under the joint density function $p(\omega, \kappa) \propto \omega^{\kappa - 1}(1 - \omega)^{n_\ab - 1} \kappa^{2K} e^{-b\kappa}$, $\omega \in (0,1)$, $\kappa > 0$. Consequentially, a valid Gibbs update of $\kappa$ is obtained by the following two steps:
\benum
\item Draw $\omega \sim p(\omega | \kappa)  = \bet(\omega | \kappa, n_\ab)$, and, then
\item Draw $\kappa \sim p(\kappa | \omega) \propto \omega^{\kappa} \kappa^{2K} e^{-b\kappa} = \gam(\kappa | 2K + 1, b - \log(\omega))$
\eenum

\paragraph{Step 5. Other updates.} Following the recommendation of \cite{neal2000markov}, we carry out additional Gibbs like updates for the cluster specific parameters $(\phi^*_c, \psi^*_c, \bfpi^*_c)$, $c = 1, \ldots, K$, to improve mixing of the Markov chain sampler. For any cluster $c$, let $S_c = \{1 \le j  \le n_\ab : (\phi_j, \psi_j, \bfpi_j) = (\phi^*_c, \psi^*_c, \bfpi^*_c)\}$ denote the collection of AB trials belonging to that cluster. The conditional posterior distribution of $(\phi^*_c, \psi^*_c, \bfpi^*_c)$, given all other model parameters and latent variables, then depend only on the reduced model:
\[
\bfeta_j \sim \nm(\phi^*_c {\bf 1}, \psi^*_c \bfC_{\ell_j}),~~\ell_j \sim P_{\scL, \bfpi^*_c}, ~j \in S_c,~~~(\phi^*_c, \psi^*_c, \bfpi^*_c) \sim G_\kappa.
\]
First, by utilizing the multinomial-Dirichlet conjugacy, we update $\bfpi^*_c$ by making a draw from $\dir(a_1 + n^c_1, \ldots, a_L + n^c_L)$ where $n^c_i = \#\{j \in S_c: \ell_j = \ell^*_i\}$, $i = 1, \ldots, L$. Next, calculate four {\it summary statistics} for the cluster:
\[
u_c = \sum_{j \in S_c} {\bf1}^T \bfC_{\ell_j}^{-1}{\bf1},~~v_c = \sum_{j \in S_c} {\bf1}^T \bfC_{\ell_j}^{-1}{\bfeta_j},~~w_c = \sum_{j \in S_c} {\bfeta_j}^T \bfC_{\ell_j}^{-1}\bfeta_j,~\mbox{and, }z_c = \frac{v_c}{u_c} 
.\]
From the reduced model and the choice of $G_\kappa$, we write $p(\phi^*_c, \psi^*_c | -) =  p(\psi^*_c | -)p(\phi^*_c | \psi^*_c, -)$ where
\beq
p(\phi^*_c | \psi^*_c, -) \propto \nm(\phi^*_c | 0, 1 - \psi^*_c) \prod_{j \in S_c} \nm(\bfeta_j | \phi^*_c{\bf 1}, \psi^*_c \bfC_{\ell_j}) \propto \nm(\phi^*_c | m_c, s_c^2) 
\label{phi-normal}
\eeq
with $s_c^2 = \psi^*_c (1 - \psi^*_c)/\{\psi^*_c + (1 - \psi^*_c)u_c\}$, $m_c = (1 - \psi^*_c )v_c/\{\psi^*_c + (1 - \psi^*_c)u_c\}$, and,
\begin{align}
p(\psi^*_c  & | -) \propto \bet(\psi^*_c | 1, \kappa) \int \nm(\phi^*_c |0, 1 - \psi^*_c) \prod_{j \in S_c} \nm(\bfeta_j | \phi^*_c{\bf 1}, \psi^*_c \bfC_{\ell_j})d\phi^*_c \nonumber \\
& \propto \bet(\psi^*_c | 2, \kappa) \nm\left(z_c | 0, {\psi^*_c}{u_c^{-1}} + 1 - \psi^*_c\right) \textit{IG}\left(\psi^*_c \left| \tfrac{M|S_c| - 1}{2}, \tfrac{w_c - z_c^2u_c}2\right.\right). \label{invgam}
\end{align}
The shape parameter of the inverse-gamma distribution on the last expression is well defined whenever the cluster size $|S_c|$ is at least two. The rate parameter is always well defined by the Cauchy-Schwartz theorem since $\langle(\bfa_j : j \in S_c), (\bfb_j : j \in S_c)\rangle = \sum_{j \in S_c} \bfa_j^T \bfC_{\ell_j}^{-1} \bfb_j$ defines an inner product on $(\bbR^{M})^{S_c}$. 

Consequently, we update $\psi^*_c$ by a Metropolis-Hastings step where we propose $\psi'_c$ from the inverse-gamma distribution in \eqref{invgam}, and, accept the proposal with probability given by the Hastings ratio $\min\{1, f(\psi'_c)/f(\psi^*_c)\}$ where $f(\psi) = \bet(\psi | 2, \kappa) \nm(z_c | 0, \psi u_c^{-1} + 1 - \psi)$. Next $\phi^*_c$ is updated by making a draw from the normal distribution on the extreme right of \eqref{phi-normal}.

\section{Dirichlet process prior on $\bbQ$}
An alternative prior specification on $\bbQ$ is to directly assign it a Dirichlet process prior distribution. This would amount to modeling $\bbQ$ as
\[
\bbQ = \sum_{h = 1}^\infty \omega_h \delta_{(\phi_h^*, \psi^*_h, \ell^*_h)}
\]
where the atoms $(\phi^*_h, \psi^*_h, \ell^*_h)$, $h = 1, 2, \ldots$, are drawn independently from a base measure $G'_\kappa$ on $(-\infty, \infty) \times (0, \infty) \times \scL$, and, the weights $\omega_h$, $h = 1, 2, \ldots$, are given by the stick-breaking construction as detailed in Section \ref{dp prior}. To mimic the specification of Section \ref{base m}, the base measure $G'_\kappa$ could be taken as in \eqref{g0}, but with the part $\bfpi^* \sim \dir(a_1, \ldots, a_K)$ now replaced with a matching counterpart $\ell^* \sim \scP_{\scL^*, \bar a}$ where $\bar a$ gives the probability vector obtained by normalizing $a = (a_1, \ldots, a_K)$. 

In contrast to the DAPP specification, this alternative formulation enforces a hard coupling between $\ell$ and $(\phi, \psi)$ by removing the intermediary $\bfpi$ in \eqref{bfpi}. Here, all \ab~trials in a cluster must share a common value for each of these three parameters. While this specification appears simpler than ours, and, a hard coupling may be scientifically meaningful, it leads to more challenging posterior computation. Specifically, in our experience, an adaptation of Algorithm \ref{algo1} to the alternative formulation, leads to considerably poorer mixing of the resulting Markov chain sampling.

Figure \ref{fig alt} compares Monte Carlo estimates of the posterior predictive distribution of $\ell$ from three independent runs of the MCMC in fitting the DAPP model or its alternative to the synthetic data set from Experiment 3 (Section \ref{case studies}). Each Markov chain was run for 10,000 iterations of which first 1,000 draws were discarded, and, 1,000 samples were saved from the remainder of the chain by thinning it uniformly. We measured Monte Carlo error as $\max_c \|p^c - \bar p\|_1$, where $p^c$ denotes the posterior predictive probability vector from chain $c$, $c \in \{1, 2, 3\}$ and $\bar p$ is the average across the three chains. For the DAPP model, the Monte Carlo error is 0.07 whereas for the alternative model it equals 0.37.

\begin{figure}[ht]
\graphsc{0.6}{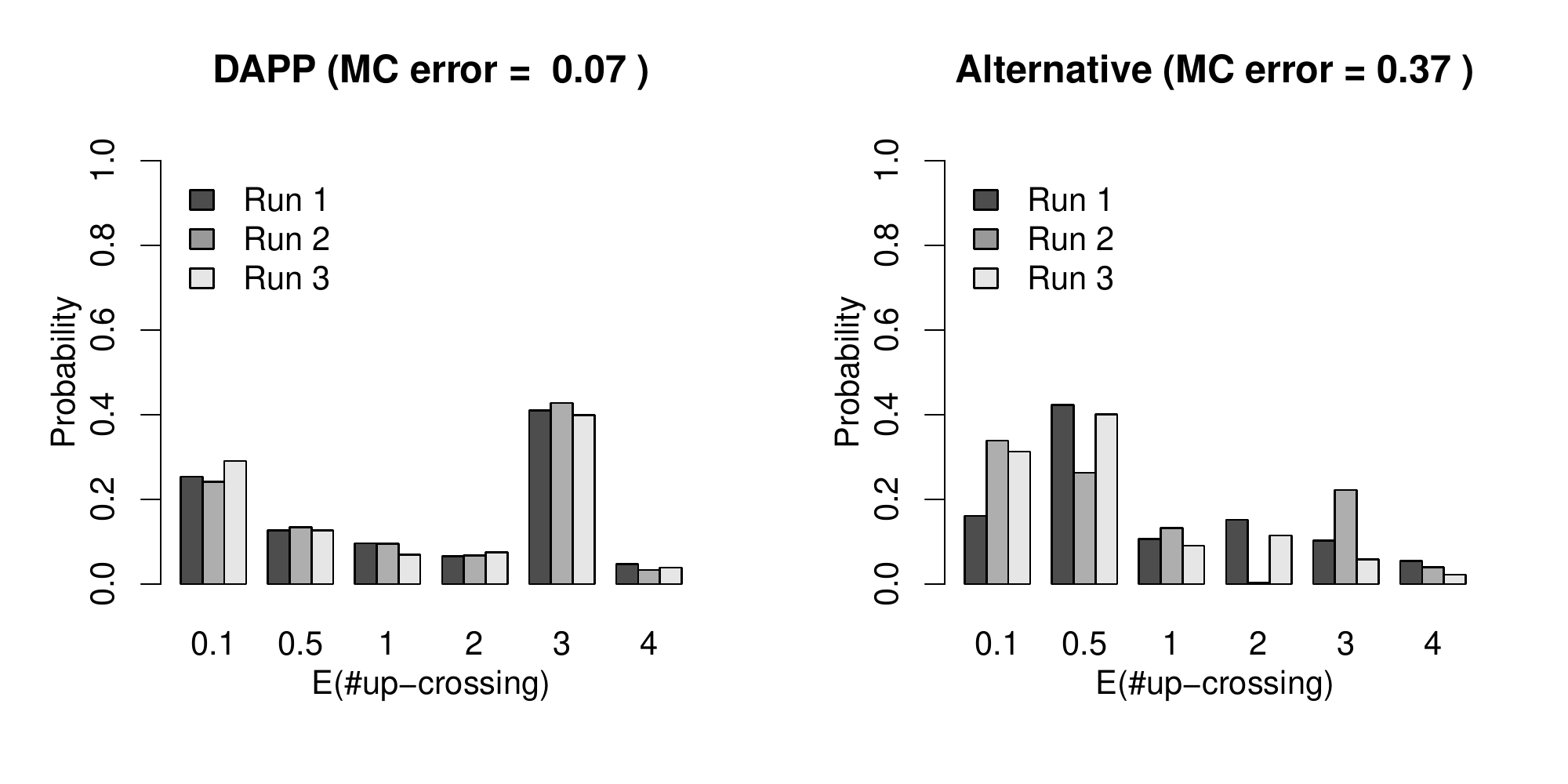}
\caption{Evidence of poor MCMC mixing when a Dirichlet process prior is assigned directly to $\bbQ$, as opposed to the hierarchical formulation adopted in DAPP. Left panel shows estimates of the posterior predictive distribution of $\ell$ from three independent MCMC runs of DAPP. Right panel  shows the same for the alternative specification. The data set from Experiment 3 is used for model fitting.}
\label{fig alt}
\efig

\end{appendices}

\bibliographystyle{chicago}
\bibliography{MetaBib.bib}

\end{document}

%% file: mycommands.tex
\usepackage{amssymb,amsmath,amsthm,graphicx}
\usepackage{textpos, url, multicol, multirow}
\usepackage{wrapfig, rotating}
\usepackage[usenames,dvipsnames]{color}
\usepackage{hyperref}
\hypersetup{
 colorlinks,
 citecolor=Black,
 linkcolor=Black,
 urlcolor=Blue}

\newcommand{\bs}{\begin{slide}}
\newcommand{\es}{\end{slide}}

\newcommand{\bitem}{\begin{itemize}}
\newcommand{\eitem}{\end{itemize}}

\newcommand{\benum}{\begin{enumerate}}
\newcommand{\eenum}{\end{enumerate}}

\newcommand{\beq}{\begin{equation}}
\newcommand{\eeq}{\end{equation}}

\newcommand{\beqr}{\begin{eqnarray*}}
\newcommand{\eeqr}{\end{eqnarray*}}

\newcommand{\beqrn}{\begin{eqnarray}}
\newcommand{\eeqrn}{\end{eqnarray}}

\newcommand{\bfig}{\begin{figure}
\centering}
\newcommand{\efig}{\end{figure}}

\newcommand{\graphsc}[2]{\includegraphics[scale = #1]{#2}}

\newcommand{\bfa}{\mathbf{a}}
\newcommand{\bfb}{\mathbf{b}}

\newcommand{\bfm}{\mathbf{m}}

\newcommand{\bfp}{\mathbf{p}}

\newcommand{\bfy}{\mathbf{y}}

\newcommand{\bfC}{\mathbf{C}}

\newcommand{\bfS}{\mathbf{S}}

\newcommand{\bfX}{\mathbf{X}}
\newcommand{\bfY}{\mathbf{Y}}
\newcommand{\bfZ}{\mathbf{Z}}

\newcommand{\bfeta}{\boldsymbol{\eta}}

\newcommand{\bfpi}{\boldsymbol{\pi}}

\newcommand{\bftheta}{\boldsymbol{\theta}}

\newcommand{\scL}{\mathcal{L}}

\newcommand{\scP}{\mathcal{P}}

\newcommand{\bbP}{\mathbb{P}}
\newcommand{\bbQ}{\mathbb{Q}}
\newcommand{\bbR}{\mathbb{R}}

\newcommand{\expct}{\mathbb{E}}

\newcommand{\cov}{\mathbb{C}\mbox{ov}}

\newcommand{\bet}{\textit{Be}}
\newcommand{\bin}{\textit{Bin}}

\newcommand{\dir}{\textit{Dir}}
\newcommand{\dpp}{\mathrm{DP}}

\newcommand{\gam}{\textit{Ga}}

\newcommand{\gp}{\mathrm{GP}}

\newcommand{\nm}{\textit{N}}

\newcommand{\pois}{\textit{Poi}}

\newcommand{\diag}{\mathrm{diag}}

\theoremstyle{plain}

\theoremstyle{definition}

\theoremstyle{remark}
\newtheorem{rem}{Remark}

\newtheorem{tasktext}{\sc Task}

\newtheorem{restext}{\sc Result}